\documentclass[showpacs,aps,preprintnumbers,prb,amsmath,amssymb,twocolumn,floatfix]{revtex4-1}
\usepackage{graphicx}
\usepackage{color}
\usepackage[normalem]{ulem}
\usepackage{amssymb}
\usepackage{amsmath}
\usepackage[normalem]{ulem}
\DeclareGraphicsExtensions{.pdf,.eps,.png,.jpg,.mps}

\begin{document}

\title{The effective increase in atomic scale disorder by doping \\ and superconductivity in Ca$_3$Rh$_4$Sn$_{13}$. }
\author{A.~\'{S}lebarski$^{1,2}$, P.~Zajdel$^{1}$, M.~Fija\l kowski$^{1}$, M.~M.~Ma\'{s}ka$^{1}$,  P.~Witas$^{1}$, J.~Goraus$^{1}$,  Y.~Fang$^{3,4}$, D.~C.~Arnold$^{5}$, and M.~B.~Maple$^{3}$}
\affiliation{
$^{1}$Institute of Physics,
University of Silesia in Katowice, Uniwersytecka 4, 40-007 Katowice, Poland\\
$^{2}$Centre for Advanced Materials and Smart Structures, 
Polish Academy of Sciences, Ok\'{o}lna 2, 50-950 Wroc\l aw, Poland\\
$^{3}$Department of Physics, University of California, San Diego, La Jolla,
California 92093, USA\\
$^{4}$Materials Science and Engineering Program, University of California, San Diego, La Jolla, California 92093, USA\\
$^{5}$School of Physical Sciences, University of Kent, Canterbury, Kent CT2 7NH, UK\\ }
\begin{abstract}

The comprehensive research of the electronic structure, thermodynamic and electrical transport  properties reveals the existence of inhomogeneous superconductivity due to structural disorder in Ca$_3$Rh$_4$Sn$_{13}$ doped with La (Ca$_{3-x}$La$_x$Rh$_4$Sn$_{13}$) or Ce (Ca$_{3-x}$Ce$_x$Rh$_4$Sn$_{13}$) with superconducting critical temperatures $T_c^{\star}$ higher than those ($T_c$) observed in the parent compounds. The $T-x$ diagrams and the entropy $S(x)_T$ isotherms well document the relation between  degree of an atomic disorder and separation of the  {\it high-temperature} $T_c^{\star}$  and  $T_c$-bulk phases. In these dirty superconductors with the mean free path much smaller than the coherence length, the Werthamer-Helfand-Hohenber theoretical model does not well fits the $H_{c2}(T)$ data. We suggest that this can result from two-band superconductivity or from the presence
of strong inhomogeneity in these systems. The multiband model very well describes the 
$H-T$ dependencies, but the present results as well as our previous studies give 
arguments for the scenario based on the presence of nanoscopic inhomogeneity of the superconducting state. We also revisited the nature of structural phase transition at $T^{\star}\sim 130-170$ K and documented that there might be another precursor transition at higher  temperatures. The impact of the magnetic Ce-Ce correlations  on the increase of $T_c$ in respect to the critical temperatures of Ca$_{3-x}$La$_x$Rh$_4$Sn$_{13}$ is also discussed.

\end{abstract}


\maketitle

\section{Introduction}

The family of $R_3M_4$Sn$_{13}$ compounds, where R is an alkali metal or rare earth and $M$ is a transition metal (Ir, Rh, Ru, or Co), was first synthesized by Remeika {\it et al.} \cite{Remeika80}. Recently, there has been a resurgence of interest among the condensed matter community due to unusual properties of these materials, characterized by strong electron correlation effects \cite{Slebarski2015a}, 
structural phase transitions associated with the Fermi surface reconstruction \cite{Klintberg2012,Tompsett2014,Kuo2015,Chen2016}, and superconductivity \cite{Israel2005,Goh2015,Cheung2016}. 
Ca$_{3}$Rh$_{4}$Sn$_{13}$, a member of this skutterudite-related family is a good model material to study the various low-temperature and structural properties. Ca$_{3}$Rh$_{4}$Sn$_{13}$ adopts the P{\it m}$\bar{3}${\it n} cubic structure and has been found to be a BCS superconductor with a superconducting transition temperature $T_c$ of about 8.4 K, which can be strongly reduced by antisite defects generated by different heat treatment \cite{Westerveld1987,Westerveld1989}. 
Similarly, an atomic disorder can occur as a result of doping.
Recently, we documented experimentally,  that the  effect of nanoscale disorder generated by doping of the Ca$_{3}$Rh$_{4}$Sn$_{13}$~ \cite{Slebarski2016} and isostructural La$_{3}M_{4}$Sn$_{13}$~  \cite{Slebarski2014a} superconductors
leads to appearance of an inhomogeneous superconducting state, characterized by the critical temperature $T^{\star}_c$ higher than  $T_c$ of the bulk phase. Similar interesting behavior has been observed in a number of other strongly-correlated superconductors (see, e.g., \cite{Maple2002,Vollmer2003,Bianchi01,Seyfarth2006,Meason2008,Bianchi01,Slebarski2014a,Slebarski2015}), particularly those close to a quantum critical point (QCP), where an increase of $T_c$ was documented by nanoscale electronic disorder. In the critical regime, such a system is  at the threshold of an instability, and even a weak perturbation, such as disorder can cause significant macroscopic effects.
This is a reason for continuing our research of 
the atomic scale disorder and its impact on a novel phenomena in Ca$_{3}$Rh$_{4}$Sn$_{13}$ and similar materials. Moreover, for a series of R$_{3}M_{4}$Sn$_{13}$ it was claimed that the cubic crystallographic structure P{\it m}$\bar{3}${\it n} is modulated below temperature  $T^{\star}\sim 130-170$ K with a k-star of a propagation vector {\bf q}=($\frac{1}{2},\frac{1}{2},0$). The structural second order-type transition at {\it $T^{\star}$} converts the simple cubic high-temperature structure P{\it m}$\bar{3}${\it n}
into a body centered cubic structure I$4_{1}$32~ \cite{Bordet1991} with twice the lattice parameters due to the distortion of the Sn1Sn2$_{12}$ icosahedra related to a charge transfer from Sn2 toward Sn1 atoms \cite{comment2}. However, no signature of this anomaly associated with $T^{\star}$ was observed for undoped  Ca$_{3}$Rh$_{4}$Sn$_{13}$. We documented, that Ca, when is partially replaced by La (Ca$_{3-x}$La$_x$Rh$_{4}$Sn$_{13}$) or Ce (Ca$_{3-x}$Ce$_x$Rh$_{4}$Sn$_{13}$) which simulates a negative chemical pressure, revealed the existence of this structural transformation at the presence of $T_c$ and $T^{\star}_c$ superconducting phases. Recently, it was also shown that this phase transition remains second-order at $T=0$, which leads to a novel structural QCP \cite{Klintberg2012,Goh2015,Cheung2016}. In this manuscript, we discuss the impact of the magnetic correlations on the increase of $T_c$ and $T^{\star}_c$ of the Ce-doped alloys with respect to superconducting temperatures of Ca$_{3-x}$La$_x$Rh$_{4}$Sn$_{13}$. On the basis of electrical transport, thermodynamic properties, and band structure calculations we propose a phenomenological model, which qualitatively interprets the experimental data. Finally, we  revisit the effect of structural instability at $T^{\star}$ and show that there might be another, precursor transition at higher temperatures.

\section{Experimental details}   

The Ca$_{3}$Rh$_{4}$Sn$_{13}$, La$_{3}$Rh$_{4}$Sn$_{13}$ and Ce$_{3}$Rh$_{4}$Sn$_{13}$ polycrystalline samples were prepared by arc melting the constituent elements on a water cooled copper hearth in a high-purity argon atmosphere with an Al getter. 
The Ca$_{3-x}$La$_x$Rh$_4$Sn$_{13}$ and Ca$_{3-x}$Ce$_x$Rh$_4$Sn$_{13}$ alloys were then prepared by diluting the parent compounds with nominal compositions of La or Ce and Ca which were then annealed  at 870$^{o}$C for 2 weeks. All samples were examined by x-ray diffraction (XRD) analysis and in the first approximation found to have a cubic structure (space group P{\it m}$\bar{3}${\it n}) \cite{Remeika80}. \\
Variable temperature powder XRD measurements were carried out on a single crystal diffractometer Rigaku (Oxford Diffraction) Supernova in a powder mode using Cu $K_{\alpha}$ microsource (50~kV, 0.80~mA). Small amounts of samples ($< 1$ mg) were powdered before the experiments and glued to a tip of a glass rod (0.1~mm). Data were collected on a heating ramp with stops at temperatures chosen in a 90~K to 390~K range. At each temperature 2x30~s acquisitions (30 degree rotation) were collected for 8 detector positions, effectively covering 2$\theta$ range from 2 to 155 degrees. 
Synchrotron powder XRD was carried out on La$_{2.8}$Ca$_{0.2}$Rh$_{4}$Sn$_{13}$ at the Swiss-Norwegian Beamlines (SNBL) at the European Synchrotron Radiation Facility (ESRF) in Grenoble. The specimen was powdered and loaded into a 0.3~mm quartz capillary. The instrument was operated at wavelength 0.71446~{\AA} and the temperature was maintained using Cryostream 700+ temperature controller. The datasets were collected on a heating ramp with 6~K/min. The 2D images were processed using CrysAlis software package and full pattern Rietveld  refinements were carried out using Fullprof Suite\cite{fullprof}. Variable temperature ramps were merged, visualized and fit using the DAVE package\cite{DAVE}. \\
Variable temperature Raman spectra were collected on a Horiba Yvon Jobin LabRAM HR instrument using a 531~nm laser and Linkam Examina THMS 600 cold stage. Measurements were performed using twenty integrations with a 6~s acquisition time with x50 long working distance objective and 600 lines per mm grating (giving a spectral resolution of ±0.5~cm$^{-1}$) over a Raman shift range between 80~cm$^{-1}$ and 1200~cm$^{-1}$. 

The  compositions of the Ca$_{3-x}$Ce$_x$Rh$_4$Sn$_{13}$ and Ca$_{3-x}$La$_x$Rh$_4$Sn$_{13}$ samples, checked by electron microprobe technique and by XPS analysis were very close to the assumed stoichiometry. However, local fluctuations in stoichiometry over the length of the sample  were observed at the nanoscale for all $x$ components of the both systems, the greatest one exist for Ce or La, which explain the strong disorder induced by doping (c.f. the detailed investigations of the homogeneity of the series of Ca$_{3-x}$Ce$_x$Rh$_4$Sn$_{13}$ compounds are presented and discussed in Ref. \onlinecite{Slebarski2016},  similar fluctuations in the composition are observed in the system of Ca$_{3-x}$La$_x$Rh$_4$Sn$_{13}$ alloys, which signals site disorder).

\begin{figure}[h!]
\includegraphics[width=0.48\textwidth]{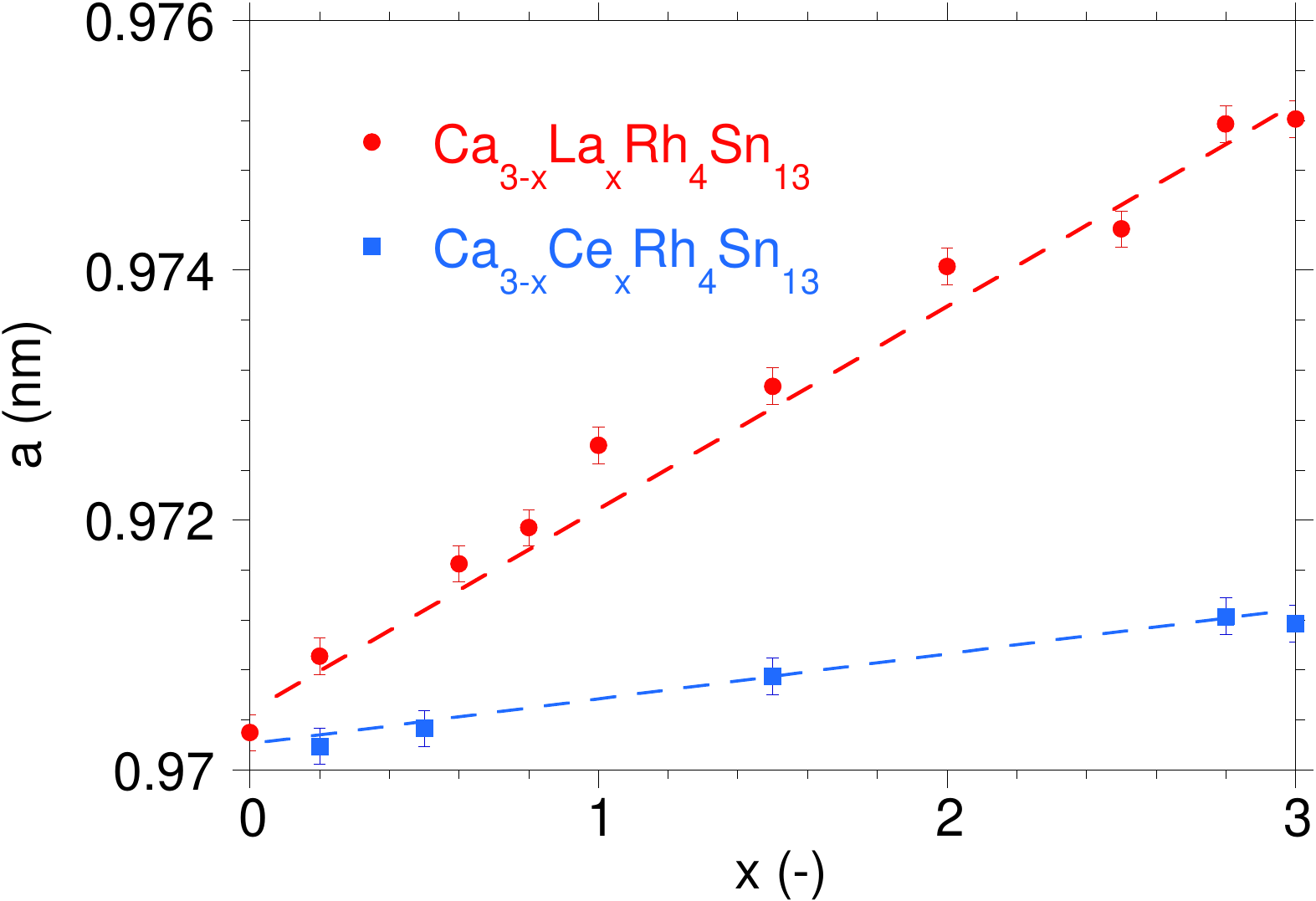}
\caption{\label{fig:Fig1_a-x-La-Ce}
The lattice parameter, $a$, plotted against La or Ce concentration, $x$, for the   Ca$_{3-x}$La$_x$Rh$_4$Sn$_{13}$ and Ca$_{3-x}$Ce$_x$Rh$_4$Sn$_{13}$ series of alloys. The lattice parameters follow Vegard's law, which suggests good sample quality and stoichiometry of the components $x$. 
}

\end{figure}

Figure \ref{fig:Fig1_a-x-La-Ce} displays the lattice parameters $a$ vs $x$ obtained at room temperature for  Ca$_{3-x}$La$_x$Rh$_4$Sn$_{13}$ and Ca$_{3-x}$Ce$_x$Rh$_4$Sn$_{13}$ samples, with an error bar determined by the experimental accuracy of $\Delta \theta =0.005^o$ for each XRD pattern. For both cases $a$ increases linearly with the increasing concentration of the dopant, although the La and Ce atomic radii are smaller than the Ca atomic radius. 
This behavior can be explained by different ionic radius of Ca$^{2+} \cong 1$ \AA, La$^{3+} \cong 1.15$ \AA, and Ce$^{3+} \cong 1.11$ \AA, respectively, which suggests the localized character of $f$-electron bands and the localized magnetic moment of Ce.
\indent Electrical resistivity $\rho$ at ambient pressure and magnetic fields up to 9 T was investigated by a conventional four-point ac technique using a Quantum Design Physical Properties Measurement System (PPMS). 
Measurements of $\rho$ under pressure were performed in a piston-cylinder clamped cell (for details, see Ref. \onlinecite{Slebarski2012,Fang2015}).\\ 
\indent Specific heat $C$ was measured in the temperature range $0.4-300$ K and in external magnetic fields up to 9 T using a Quantum Design PPMS platform. The dc magnetization $M$ and (dc and ac) magnetic susceptibility $\chi$ were obtained using a commercial superconducting quantum interference device magnetometer from 1.8 K to 300 K in magnetic fields up to 7 T.\\
\indent The XPS spectra were obtained
with monochromatized Al $K_{\alpha}$ radiation at room temperature
using a PHI 5700 ESCA spectrometer. The  sample was broken under high vacuum better than $6\times 10^{-10}$ Torr immediately before taking a spectrum.\\
\indent The refined lattice parameters shown in the Fig. \ref{fig:Fig1_a-x-La-Ce}  and corresponding atomic positions were used in our band structure calculations.  
The band structure calculations were accomplished using fully relativistic full potential local orbital method
(FPLO9-00-34 computer code \cite{fplo}) within the local spin density approximation (LSDA) as well as ELK FP-LAPW/APW+lo code \cite{elk}. The exchange correlation potential V$_{xc}$ was used in the form proposed by Perdew-Wang \cite{pw92} in both cases.
The number of k-points in the irreducible wedge of Brillouin Zone was 80.
The results obtained from both methods were accomplished for the same V$_{xc}$, and as expected were essentially the same.
The ELK-code was used for accurate calculations of the electron localization function (ELF), whereas the FPLO method was used to study the pressure effects on the electron density of states (DOS) of the samples.

\section{Results and discussion}

\subsection{Superconductivity in the presence of disorders in  Ca$_{3}$Rh$_{4}$Sn$_{13}$ doped with La and Ce: a comparative study}
\subsubsection{Electrical resistivity; the effect of magnetic field and pressure on superconductivity}

\begin{figure}[h!]
\includegraphics[width=0.48\textwidth]{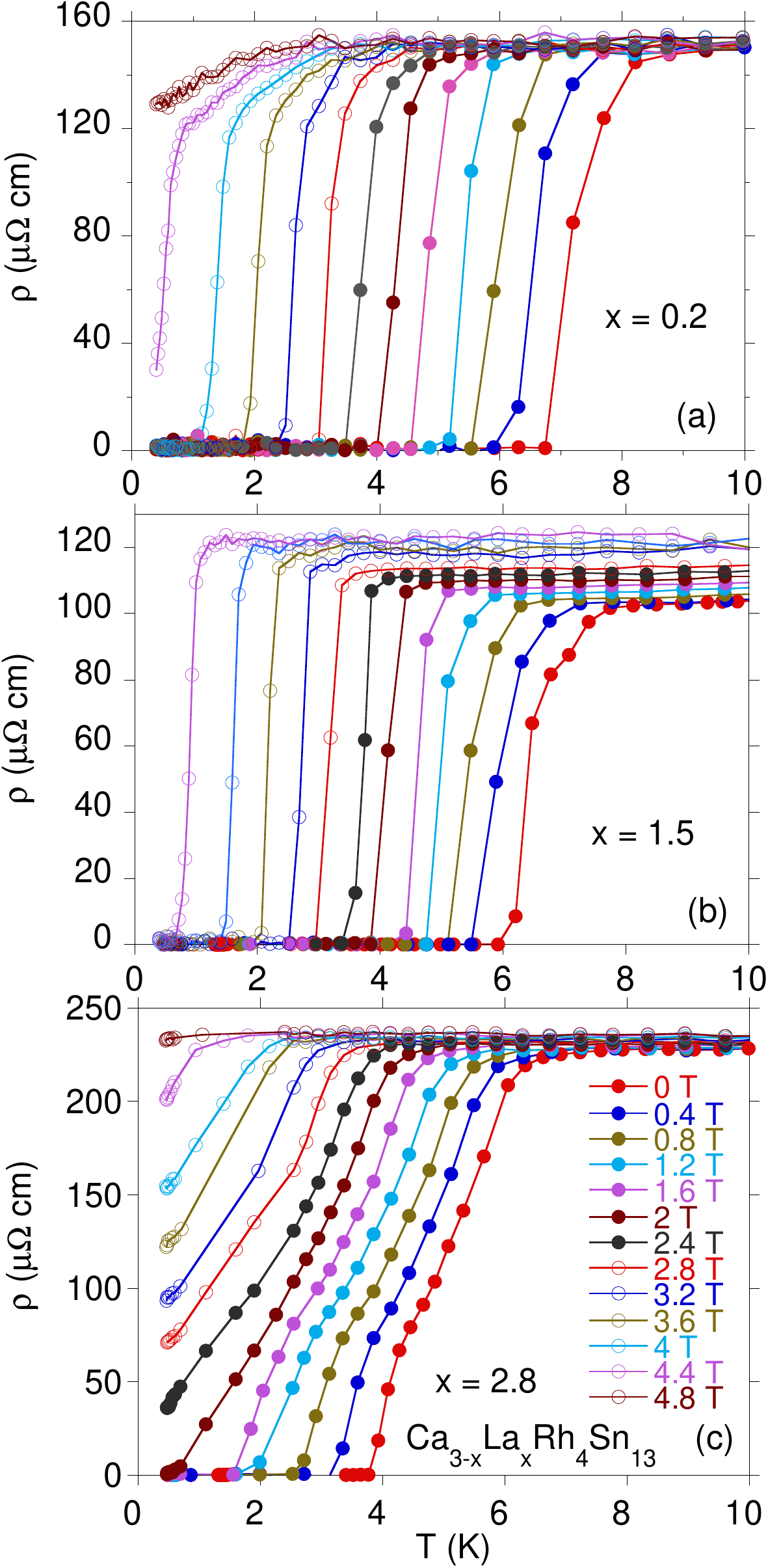}
\caption{\label{fig:Fig_R-H_sum_La-02-15-28}
Electrical resistivity for Ca$_{1-x}$La$_{x}$Rh$_4$Sn$_{13}$ ($x=0.2, 1.5$, and 2.8) at various externally applied magnetic field. The left inset shows the details near the critical temperature. The right inset displays the value of $\rho$ measured just above $T_c$ at $T=9$ K.
}
\end{figure}

We expect that an increase of crystallographic disorder by doping of Ca$_{3}$Rh$_{4}$Sn$_{13}$  will enhance the separation of the $T^{\star}_c$ and $T_c$ superconducting phases. We present a comprehensive magnetic and electrical resistivity study which indeed give evidence of these two superconducting phases.
Figure \ref{fig:Fig_R-H_sum_La-02-15-28} displays temperature  dependence of electrical resistivity $\rho (T)$ for Ca$_{3-x}$La$_x$Rh$_{4}$Sn$_{13}$ with $x=0.2$, 1.5, and 2.8 in various magnetic fields. Similar $\rho (T)$ dependencies vs $B$ were very recently reported
for the series of Ca$_{3-x}$Ce$_x$Rh$_{4}$Sn$_{13}$ alloys \cite{Slebarski2016}. The critical temperature $T^{\star}_c$ is defined as the temperature at which the resistivity falls to 50\% of its normal state value.
The transitions shown in Fig. \ref{fig:Fig_R-H_sum_La-02-15-28}
are much broader then that of Ca$_{3}$Rh$_{4}$Sn$_{13}$, which signals strong inhomogeneity due to the doping. The effect is so strong that for the alloys $x=1.5$ and 2.8, $\rho (T)$ exhibits two distinct drops, which indicate a double resistive phase transition to the superconducting state, e.g., for the sample $x=2.8$, the first resistivity drop is observed at $\sim 5.3$ K where isolated  {\it superconducting islands} begin to be formed, while the second one is at lower temperature  $T_c \sim 3.9 $ K, where a global phase coherence develops with a limit of $\rho \rightarrow 0$. This complex transition is also seen in the ac susceptibility (see section III.2.).
For Ce-substituted $x>1$ samples, a large atomic disorder may have contributed to the formation of the only inhomogeneous superconducting phase (see the $T-H$ diagram in section A.3).
Another interesting phenomenon is the observation of positive magnetoresistivity $MR = [\rho(4 T)- \rho (0)]/\rho (0)$ obvious near $T^{\star}_c$. At the critical temperature  $MR$ coefficient is about 20\% for La-doped and  about 90\% for superconducting Ce-doped alloys. The positive magnetoresistivity   can be interpreted as an effect of strong $d$-electron correlations \cite{Strydom2007,Slebarski2015a}, which dominate the field-dependent electronic transport in this {\it nonmagnetic} material \cite{comment1}.
\begin{figure}[h!]
\includegraphics[width=0.48\textwidth]{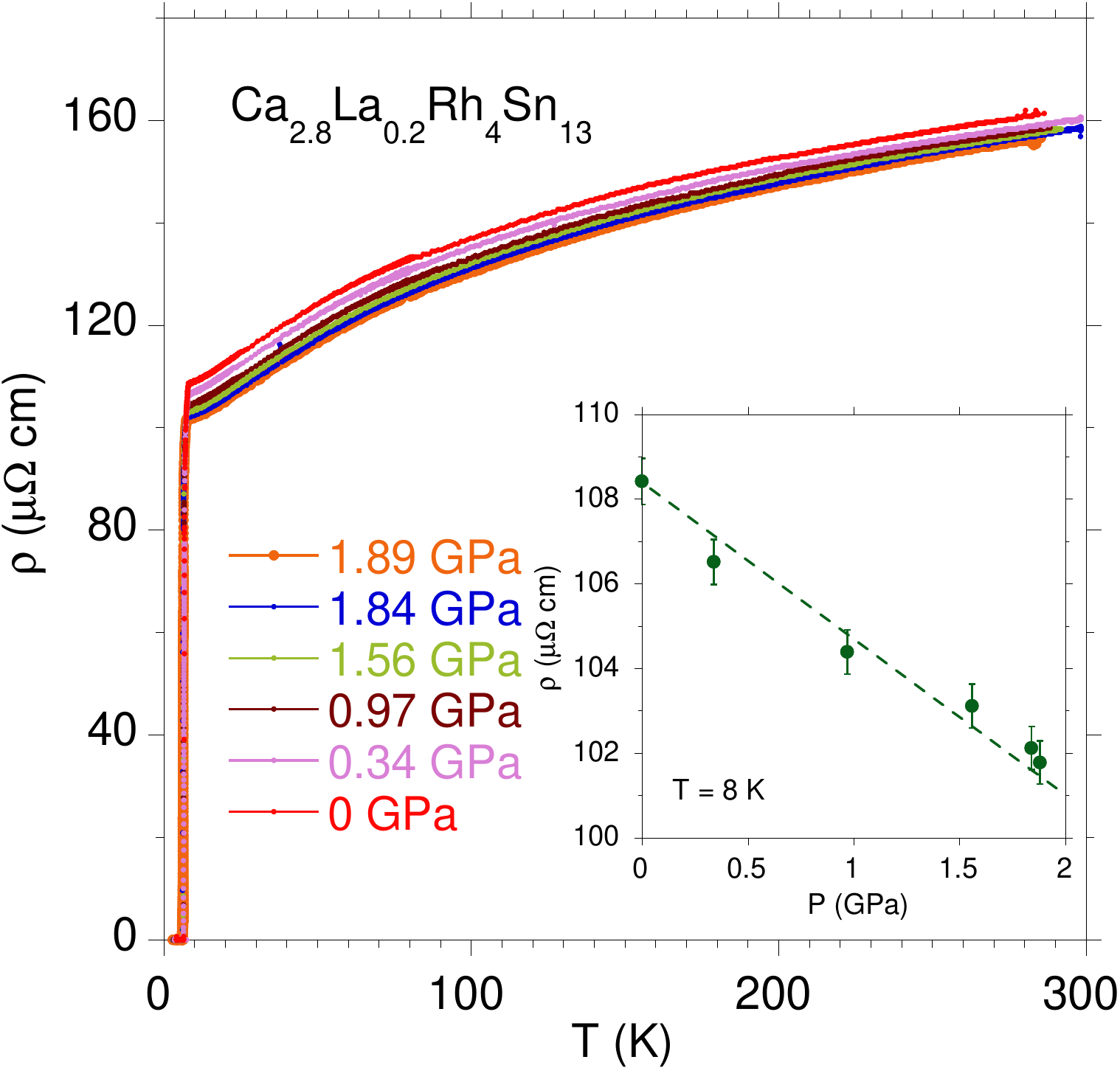}
\caption{\label{fig:Fig_R-vs-P_La-02_last}
Electrical resistivity for Ca$_{2.8}$La$_{0.2}$Rh$_4$Sn$_{13}$ under applied pressure. The inset shows the value of $\rho$ measured just above $T^{\star}_c$ at $T=8$ K.
}
\end{figure}
\begin{figure}[h!]
\includegraphics[width=0.48\textwidth]{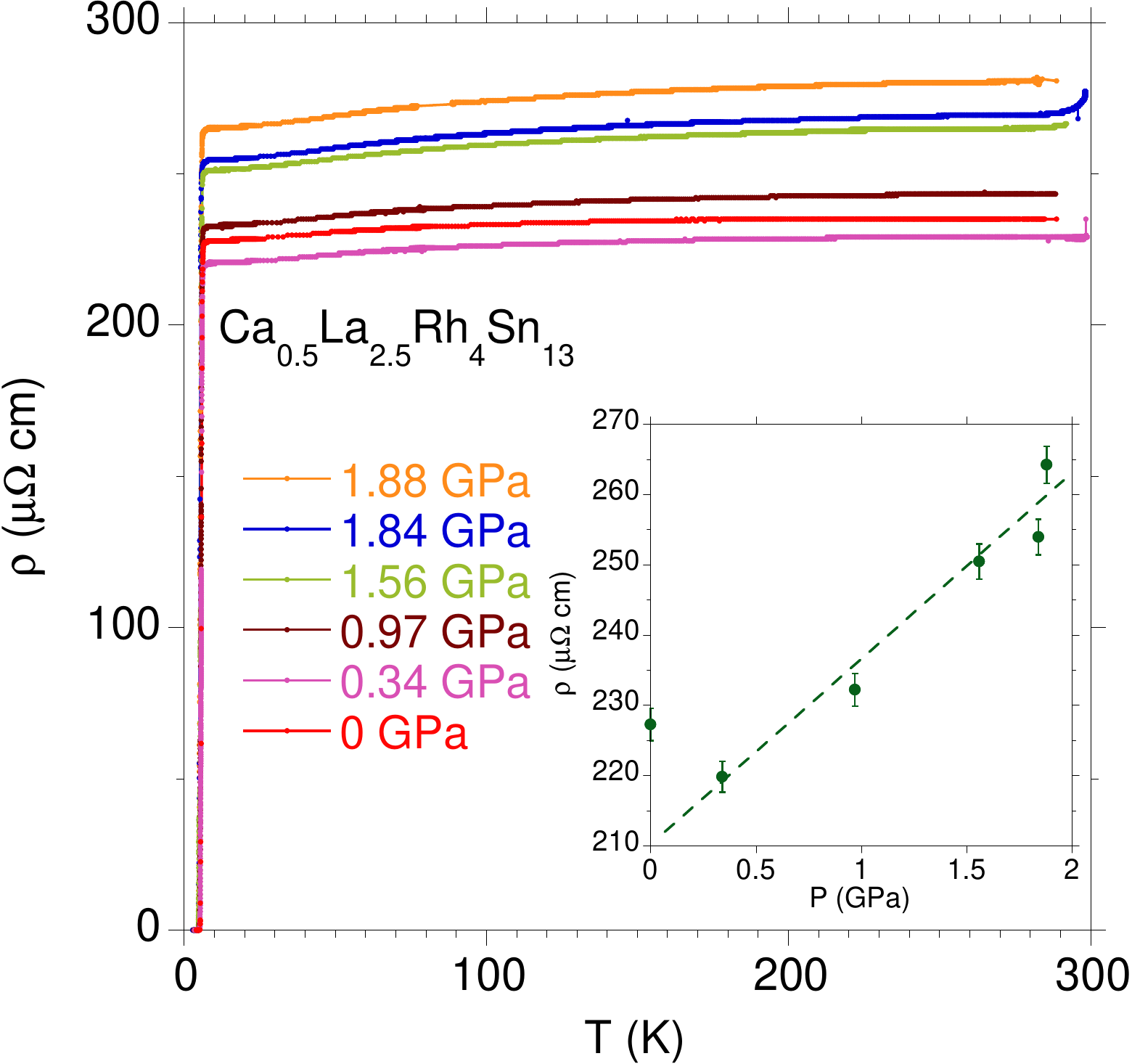}
\caption{\label{fig:Fig_R-vs-P_La-25_last}
Electrical resistivity for Ca$_{0.5}$La$_{2.5}$Rh$_4$Sn$_{13}$ under applied pressure. The inset displays the value of $\rho$ measured just above $T^{\star}_c$ at $T=7$ K.
}
\end{figure}
Figures \ref{fig:Fig_R-vs-P_La-02_last} and \ref{fig:Fig_R-vs-P_La-25_last} show   the electrical resistivity as a function of temperature for Ca$_{2.8}$La$_{0.2}$Rh$_4$Sn$_{13}$ and Ca$_{0.5}$La$_{2.5}$Rh$_4$Sn$_{13}$ under external pressure. From these data we obtained the pressure coefficients $\frac{dT_c^{\star}}{dP} = -0.19$ K/GPa for Ca$_{2.8}$La$_{0.2}$Rh$_4$Sn$_{13}$
and $-0.21$ K/GPa for Ca$_{0.5}$La$_{2.5}$Rh$_4$Sn$_{13}$, respectively. Very similar pressure coefficients of $T^{\star}_c$ are: $-0.2$ K/GPa for Ca$_{3}$Rh$_{4}$Sn$_{13}$ and $-0.3$ K/GPa for the $x=0.2$ cerium doped sample \cite{Slebarski2016}. These coefficients  $\frac{dT_c^{\star}}{dP}$ are significantly larger then the pressure coefficients of $T_c$, found in similar isostructural La-based superconductors \cite{Slebarski2015}, e.g., $\frac{dT_c}{dP}$ is only $-0.05$ K/GPa for La$_{3}$Rh$_{4}$Sn$_{13}$ \cite{Slebarski2014a}. The $P$-dependence of $T_c$ can be interpreted  according to the Eliashberg theory of superconductivity \cite{Eliashberg61} and  
the  McMillan expression, \cite{McMillan,Dynes72} 
\begin{equation}
T_c=\frac{\theta_\mathrm{D}}{1.45} \exp \left\{ \frac{-1.04(1+\lambda)}{\lambda - \mu^*(1+0.62\lambda)} \right\},              
\end{equation}
as a solution to the finite-temperature Eliashberg equations, where $\lambda$ is the electron-phonon coupling parameter, and the Coulomb repulsion $\mu^{\star}$ is assumed to be $\sim 0.1$ which is  a typical value known for $s$ and $p$ band superconductors. Our estimation gives  $\lambda \approx 0.62$ for $T_c$ phase of Ca$_{3}$Rh$_{4}$Sn$_{13}$,
and slightly higher value of $\lambda^{\star} \approx 0.63$ for its inhomogeneous $T^{\star}_c$ phase. For La$_{3}$Rh$_{4}$Sn$_{13}$
$\lambda \approx  0.52$, while $\lambda^{\star} \approx 0.59 $. Since the 
coupling $\lambda$ given by the expression \cite{McMillan,Hopfield}
\begin{equation}
\lambda=\frac{N(\epsilon_{\rm F})\langle I^2 \rangle}{M\langle \omega^2 \rangle},
\end{equation}
where $\langle I^2 \rangle$ is the square of the electronic matrix element of electron-phonon interactions averaged 
over the Fermi surface, $\langle \omega^2 \rangle$ is an average of the square of the phonon frequency ($\omega \sim \theta_D$), $N(\epsilon_{\rm F})$ is a density of states at the Fermi energy, and $M$ is  the atomic mass,
is larger for the inhomogeneous superconducting $T_c^{\star}$ state with respect to the bulk effect observed below $T_c$, 
the primary reason for $\frac{dT_c^{\star}}{dP} > \frac{dT_c}{dP}$ is the pressure dependence of $\theta_D$, which leads to larger lattice stiffening in the $T_c^{\star}$ phase with respect to the bulk effect below $T_c$ and contributes to the $T_c^{\star}>T_c$ effect.
The $P$ dependence of $\theta_D$ 
is given
by the Gr\"uneisen parameter $\gamma_G=-\frac{dln{\theta_D}}{dlnV}$, which determines the lattice stiffening. It was documented experimentally \cite{Shao2004} that $\gamma_G$ strongly determines the magnitude and sign of $\frac{dT_c}{dP}$. In the case of inhomogeneous superconductivity one can also suppose the dominant impact of the pressure dependence of
the DOS at the Fermi level, $\epsilon_F$, more pronounced than  in bulk superconducting phases.

\begin{figure}[h!]
\includegraphics[width=0.48\textwidth]{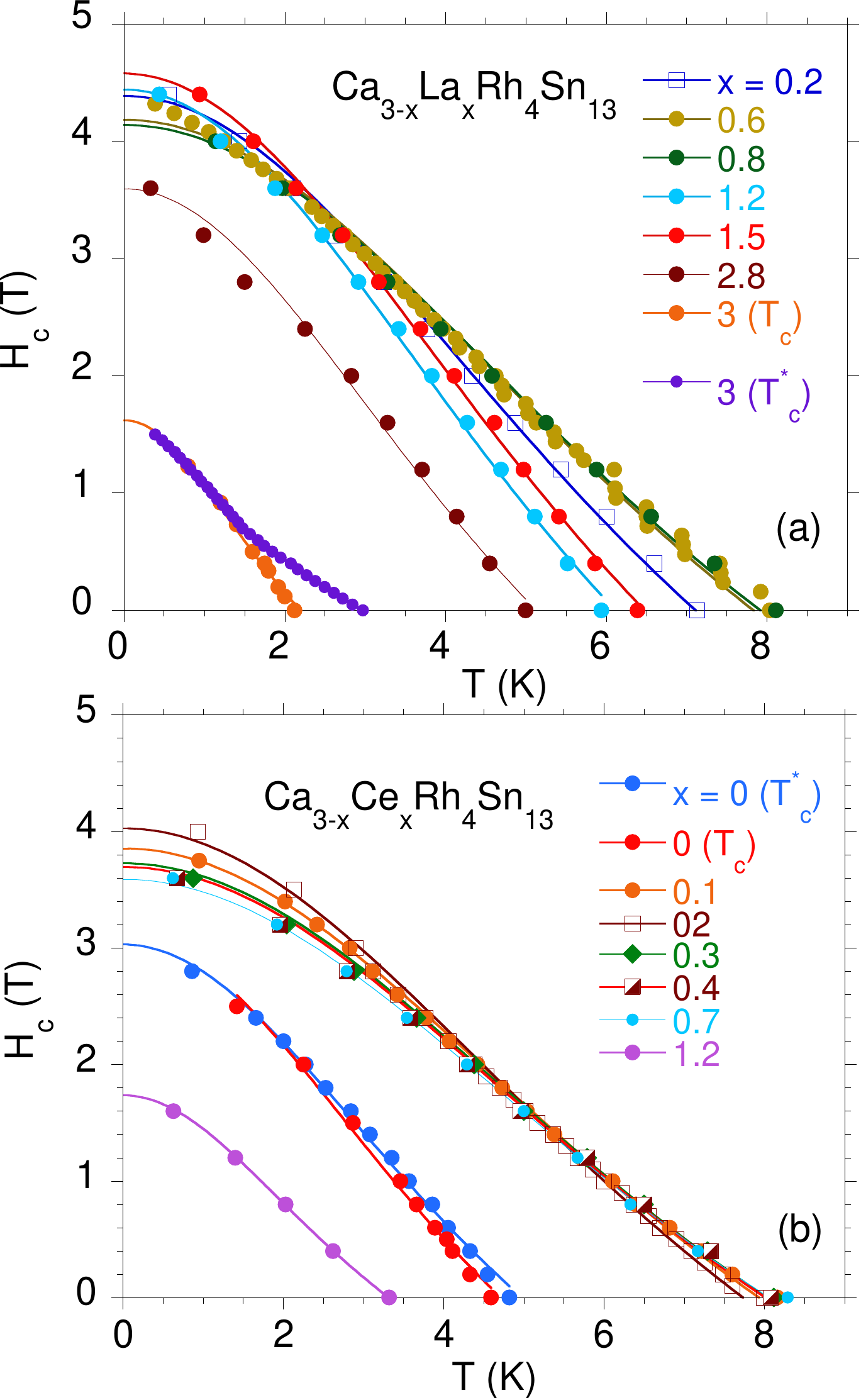}
\caption{\label{fig:Fig-DIAGRAM-H-T}
Temperature dependence of the upper critical field $H_{c2}$ and/or $H^{\star}_{c2}(0)$ in the $H-T$ phase diagram, shown for Ca$_{3-x}$La$_x$Rh$_4$Sn$_{13}$ in panel ($a$) and for  Ca$_{3-x}$Ce$_x$Rh$_4$Sn$_{13}$ in panel ($b$).
$T^{\star}_{c}$ vs $H$ data  
are obtained from electrical resistivity  under $H$, and defined as the temperature at which $\rho$ drops to 50\% of its normal-state value. $T_c$ vs $H$ data for  Ca$_{3}$Rh$_4$Sn$_{13}$ [in panel $b$] and for La$_{3}$Rh$_4$Sn$_{13}$ [in panel ($a$)] are
obtained from $C(T)/T$ vs $T$ data (see Refs. \cite{Slebarski2016,Slebarski2014a}). The solid lines represent a fit using the GL model of $H_{c2}(T)$.}
\end{figure}

Figure \ref{fig:Fig-DIAGRAM-H-T} shows the $H-T$ phase diagram of the
Ca$_{3-x}$La$_x$Rh$_{4}$Sn$_{13}$  and Ca$_{3-x}$Ce$_x$Rh$_{4}$Sn$_{13}$ alloys, respectively. 
Temperatures   $T_c^{\star}$ were obtained from the resistivity data, while  
$T_c$ were obtained from the specific heat measurements.
The Ginzburg-Landau (GL) theory fits well the data as is shown in the $H-T$ plots in Fig. \ref{fig:Fig-DIAGRAM-H-T}.
The best fit of GL equation $H_{c2}(T)=H_{c2}(0)\frac{1-t^2}{1+t^2}$, where $t=T/T_c$ gives the upper critical field values of $H_{c2}(0)$ and $H^{\star}_{c2}(0)$, where
$H^{\star}_{c2}(0)>H_{c2}(0)$, as shown in the Figure \ref{fig:Fig-DIAGRAM-H-T}. 
Moreover, a significant increase of $H^{\star}_{c2}(0) $ due to chemical doping has been documented in both lanthanum (Ca$_{3-x}$La$_{x}$Rh$_4$Sn$_{13}$) and cerium (Ca$_{3-x}$Ce$_{x}$Rh$_4$Sn$_{13}$) doped samples in respect to $H^{\star}_{c2}(0) $ of the parent compounds, e.g., $H_{c2}(0) \approx H^{\star}_{c2}(0) $  is about 3.1 T for Ca$_{3}$Rh$_4$Sn$_{13}$, 1.6 T for La$_{3}$Rh$_4$Sn$_{13}$, while $H^{\star}_{c2}(0)$ is $\sim 4.3$ T or $\sim 3.8$ T in  Ca$_{3}$Rh$_4$Sn$_{13}$ substituted with La or Ce, respectively.  Indeed, magnetization $M$ vs $B$ measurements and the residual resistivity ratio suggest an increase of $H^{\star}_{c2}(0) $ associated with a progressive change of atomic disorders.
Within the weak-coupling theory \cite{Schmidt77}, the expression 
$\mu_0 H_{c2}(0)=\frac{\Phi_0}{2\pi \xi(0)^2}$ gives
the superconducting coherence length $\xi(0)$ or $\xi^{\star}(0)$ ($\Phi_0=h/2e=2.068 \times 10^{-15}$ Tm$^2$ is the flux quantum). 
Ca$_{3}$Rh$_4$Sn$_{13}$ exhibits similar values of $\xi(0)$ and $\xi^{\star}(0)\cong 10.3 $ nm (c.f. \cite{Slebarski2016}); for La$_{3}$Rh$_4$Sn$_{13}$, $\xi(0) \approx \xi^{\star}(0)\cong 14$ nm \cite{Slebarski2014a}, while for the series of Ca$_{3-x}$La$_{x}$Rh$_4$Sn$_{13}$ and Ca$_{3-x}$Ce$_{x}$Rh$_4$Sn$_{13}$ alloys, $\xi^{\star}(0)\cong 8.6 $ nm and $\sim 9.3$ nm, respectively. 

\begin{figure}[h!]
\includegraphics[width=0.49\textwidth]{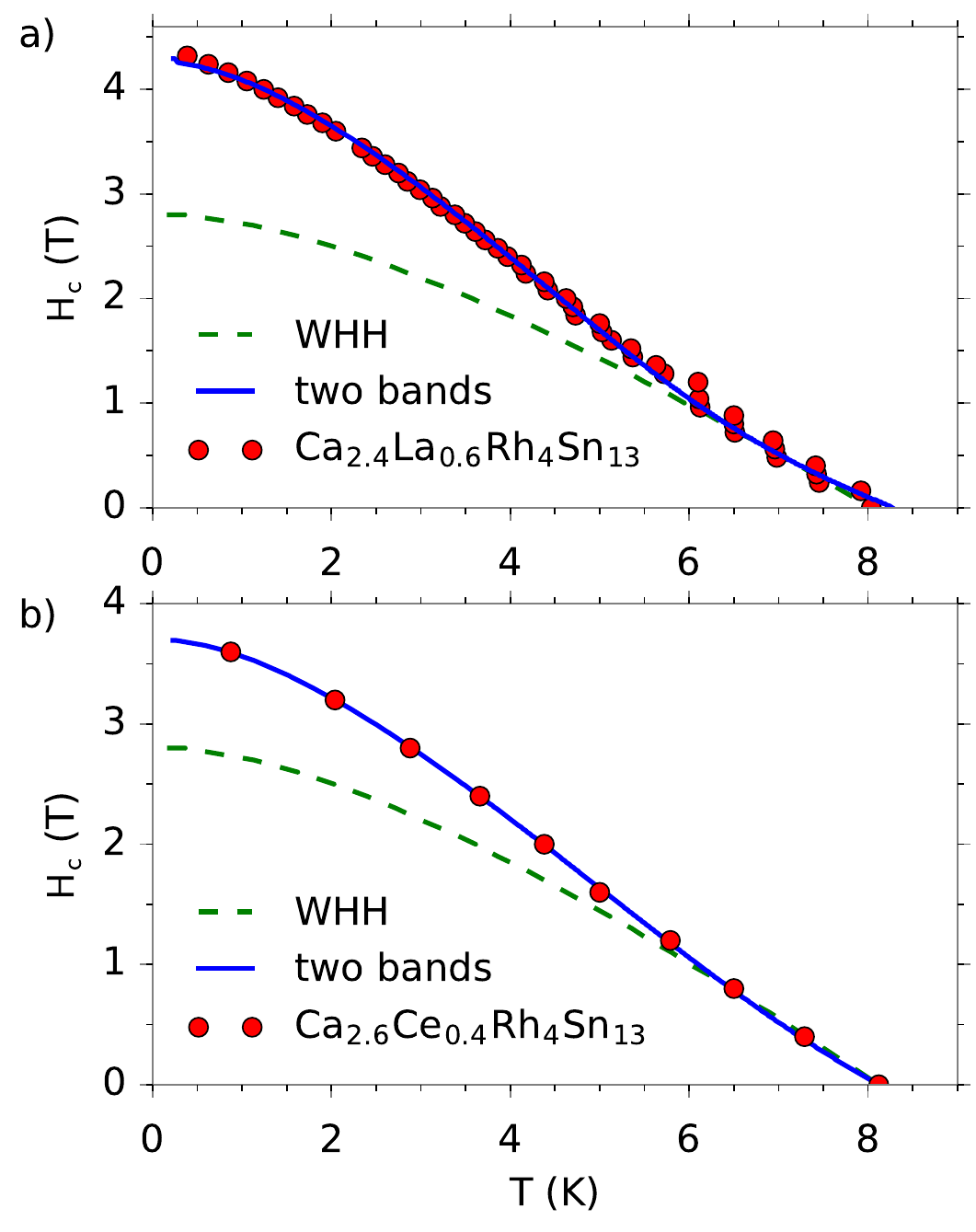}
\caption{\label{fig:hc2} Examples of the upper critical field for La-- (a) and Ce--doped (b) systems fitted by solutions of the WHH
equations. The green dashed lines marked as ``WHH'' show results of the WHH model with a single gap, whereas the solid blue lines marked as ``two bands'' represent a
two-gap model described in Ref.~\onlinecite{Gurevich}.}
\end{figure}

From the theoretical point of view, the upper critical field in a dirty superconductor, where the free mean path $l\ll \xi$,
can be described by the Werthamer-Helfand-Hohenberg (WHH) \cite{WHH} or Maki--de Gennes 
\cite{MdG} theories. The WHH theory gives 
\begin{equation}
    H_{c2}(0)=0.69\frac{dH_{c2}}{dT}T_c.
\end{equation}
It can be seen in Fig.~\ref{fig:hc2} that these
approaches underestimate $H_{c2}(T)$ at low temperatures. 
Moreover, they do not 
predict the positive curvature of $H_{c2}(T)$ close to $T_c$.

One possible explanation for this behavior of $H_{c2}(T)$ 
is multiband/multigap nature of superconductivity in these systems.
With the help of the quasiclassical Usadel equations \cite{Usadel,Usadel1} 
it was shown in Ref. \onlinecite{Gurevich} that the upper critical 
field in a two--band superconductor can be determined as a solution
of the following equation:
\begin{eqnarray}
&& a_0\left[\ln t + U(h)\right]
\left[
\ln t + U(\eta h)\right] \nonumber \\
& + & a_2\left[\ln t + U(\eta h)\right] + 
a_1\left[\ln t + U(h)\right]=0,
\end{eqnarray}
where
    $U(x)\equiv\psi\left(x+1/2\right)-\psi\left(1/2\right)$,
$\psi(\ldots)$ is the di--gamma function, $t=T/T_c$, $h$ is reduced magnetic field defined as $h=H_{c2}D_1/2\Phi_0 T$, $D_1$ 
is the band diffusivity, $\eta=D_2/D_1$. The parameters $a_{0,1,2}$
can be expressed by the intra-- and interband BCS superconducting 
coupling constants $\lambda_{11},\:\lambda_{22},\:\lambda_{12}$ and $\lambda_{21}$. The experimental data shown in Fig.~\ref{fig:hc2} can be very well reproduced by fitting these parameters within the framework of the two--band/two--gap model.
This, however, is not a solid proof of the
multiband nature of superconductivity in 
Ca$_{3-x}$La$_{x}$Rh$_4$Sn$_{13}$ and 
Ca$_{3-x}$Ce$_{x}$Rh$_4$Sn$_{13}$. One reason is that there are so many fitting parameters in the model that the agreement with the 
experimental data is relatively easy to achieve. Other  
explanations for the deviation from the WHH theory are also possible.
It is known that a positive curvature of $H_{c2}(T)$ can result
from microscopic segregation in the superconducting material, where an array of
Josephson junctions is formed\cite{Geshken,Hc2UC1,Hc2UC2,Hc2UC3}. In Refs.
\onlinecite{Slebarski2016},
\onlinecite{Slebarski2014a} and \onlinecite{Slebarski2015} we demonstrated the 
presence of two superconducting transitions due to the existence of inhomogeneous phase with  $T^{\star}$ that is different from the $T_c$ of the bulk sample. The first one corresponds to the onsets 
of inhomogeneous phase where superconductivity is present only in a fraction of the 
volume and the second one that signals the onset of bulk superconductivity. 
Arguments for this scenario can also be seen in Fig.~\ref{fig:Fig_R-H_sum_La-02-15-28}b, where a double transition, characteristic 
for inhomogeneous superconductors\cite{double-step} and suggesting the presence
of microscopic segregation, can be seen. Other explanations of the shape of $H_{c2}(T)$
are related to the presence of magnetic impurities \cite{Kresin, Kresin1, Jung2017}, 
strong 
quantization of Landau orbits \cite{Landau1,Landau2}, inhomogeneity--induced 
reduction of the diamagnetic pair--breaking \cite{Mierzejewski2002} or 
singularities in the density of states\cite{Maska2001}.

\subsubsection{Magnetic properties, evidence of short range magnetic order in Ce-doped alloys and the superconducting state}

\begin{figure}[h!]
\includegraphics[width=0.48\textwidth]{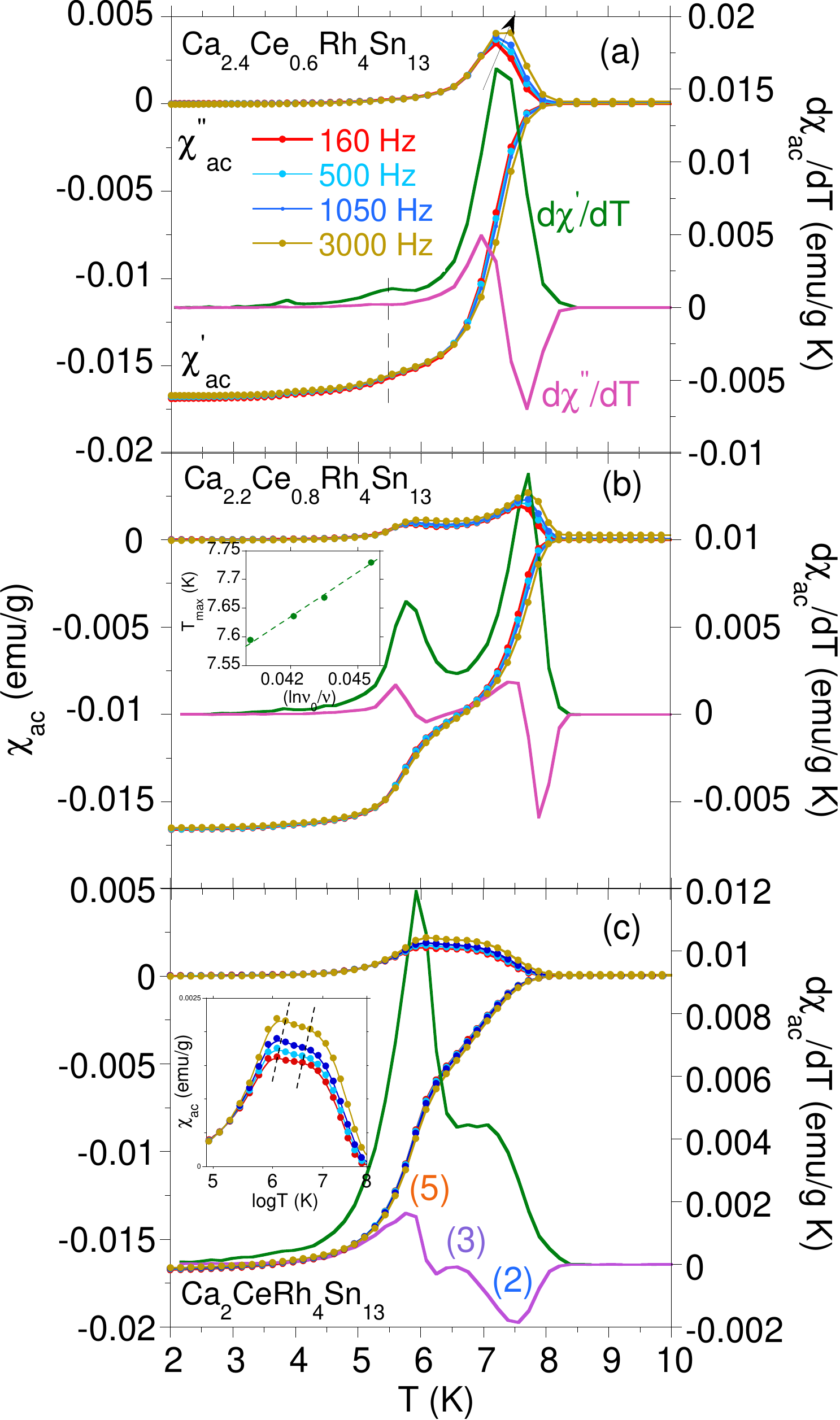}
\caption{\label{fig:Fig_CHI_ac_Ce-sum}
The real and imaginary components of the ac magnetic susceptibility, $\chi^{'}$ and $\chi^{''}$, for Ca$_{3-x}$Ce$_x$Rh$_{4}$Sn$_{13}$, as a function of temperature measured at different frequencies in a field $B=2$ Oe. The derivative $d\chi^{'}/dT$ and $d\chi^{''}/dT$ are also presented. The inset to panel ($b$) and ($c$) shows how $\chi^{''}$ depends on frequency with evidence of the spin-glass state. The minimum (2) of $d\chi^{''}/dT$ exhibits the temperature at which the inhomogeneous superconducting phase is formed,  while  the maxima (3) and (5) defines temperature $T^{\star}_c$ and $T_c$, respectively.
}
\end{figure}

\begin{figure}[h!]
\includegraphics[width=0.48\textwidth]{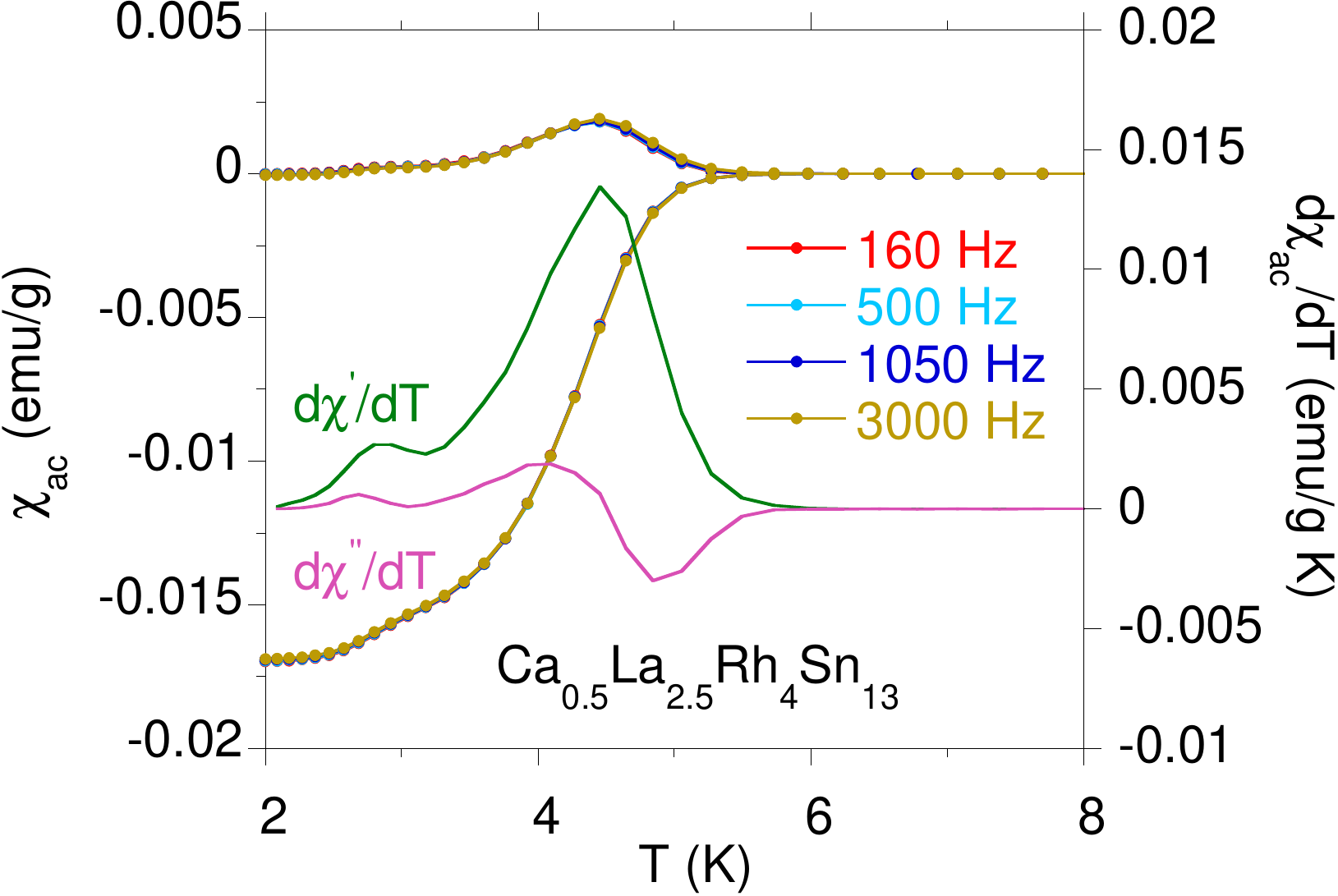}
\caption{\label{fig:Fig_CHI_ac_La-25}
The real and imaginary components of the ac magnetic susceptibility, $\chi^{'}$ and $\chi^{''}$, for Ca$_{0.5}$La$_{2.5}$Rh$_{4}$Sn$_{13}$, as a function of temperature measured at different frequencies in a field $B=2$ Oe. Also are presented the  derivative $d\chi^{'}/dT$ and $d\chi^{''}/dT$. The frequency dependence characteristic of the spin-glass-like phase is not detected for the components $x$ of Ca$_{3-x}$La$_{x}$Rh$_{4}$Sn$_{13}$. The interpretation of $d\chi^{''}/dT$ is similar to that, shown in Fig. \ref{fig:Fig_CHI_ac_Ce-sum}.
}
\end{figure}

The superconducting state of Ca$_{3}$Rh$_{4}$Sn$_{13}$ is strongly dependent on the atomic disorder, which, upon quenching, leads to a significant decrease in  $T_c$~ \cite{Westerveld1987,Westerveld1989}. 
 Our simple model explains this observation based on the assumption that the atomic disorder leads to local stress \cite{Slebarski2016}. We have calculated the systematic decrease of the density of states (DOS) at the Fermi energy  with pressure \cite{Slebarski2016} and documented 
 for Ca$_{3}$Rh$_{4}$Sn$_{13}$  obtained under various technological treatment, 
 that the DOS change well correlates with this decrease of $T_c$.
 According to this model, even a slight change in the DOS at $E_F$ may cause a significant change in $T_c$. 
With this motivation, we present a magnetic study of Ca$_{3}$Rh$_{4}$Sn$_{13}$ substituted with La and Ce to demonstrate evidence of nanoscale disorder as a bulk property, leading to an inhomogeneous superconducting state with an enhanced critical temperature $T^{\star}_c>T_c$. Here, $T^{\star}_c$ represents a drop of resistivity due to formation of percolation paths, while $T_c$ determined from magnetic susceptibility and specific heat, indicates the onset of bulk superconductivity. A comparative study has shown, that the effect of short-range magnetic correlations has a significant effect on $T_c$.
Figures \ref{fig:Fig_CHI_ac_Ce-sum} and \ref{fig:Fig_CHI_ac_La-25} compare frequency dependence of the real ($\chi^{'}$) and imaginary ($\chi^{''}$) parts of ac mass magnetic susceptibility $\chi_{ac}$, and show derivative $d\chi^{'}/dT$ and $d\chi^{''}/dT$ for the selected Ca$_{3-x}$Ce$_x$Rh$_{4}$Sn$_{13}$ samples and for Ca$_{0.5}$La$_{2.5}$Rh$_{4}$Sn$_{13}$, characteristic of the Ca$_{3-x}$La$_{x}$Rh$_{4}$Sn$_{13}$ series. 

Frequency $\nu$ dependencies in $\chi^{'}$ and $\chi^{''}$, depicted in Fig. \ref{fig:Fig_CHI_ac_Ce-sum} with characteristic Vogel-Fulcher-like behavior \cite{Mydosh} shown in the inset of panel $(b)$, become apparent  of spin-glass-like magnetic correlations in Ce doped alloys, while the $\nu$ effect is not observed for Ca$_{3}$Rh$_{4}$Sn$_{13}$ doped with La. 
The maxima in derivative $d\chi^{'}/dT$ and $d\chi^{''}/dT$ we  assigned, respectively to 
critical temperatures $T^{\star}_c$ and $T_c$.

\begin{figure}[h!]
\includegraphics[width=0.48\textwidth]{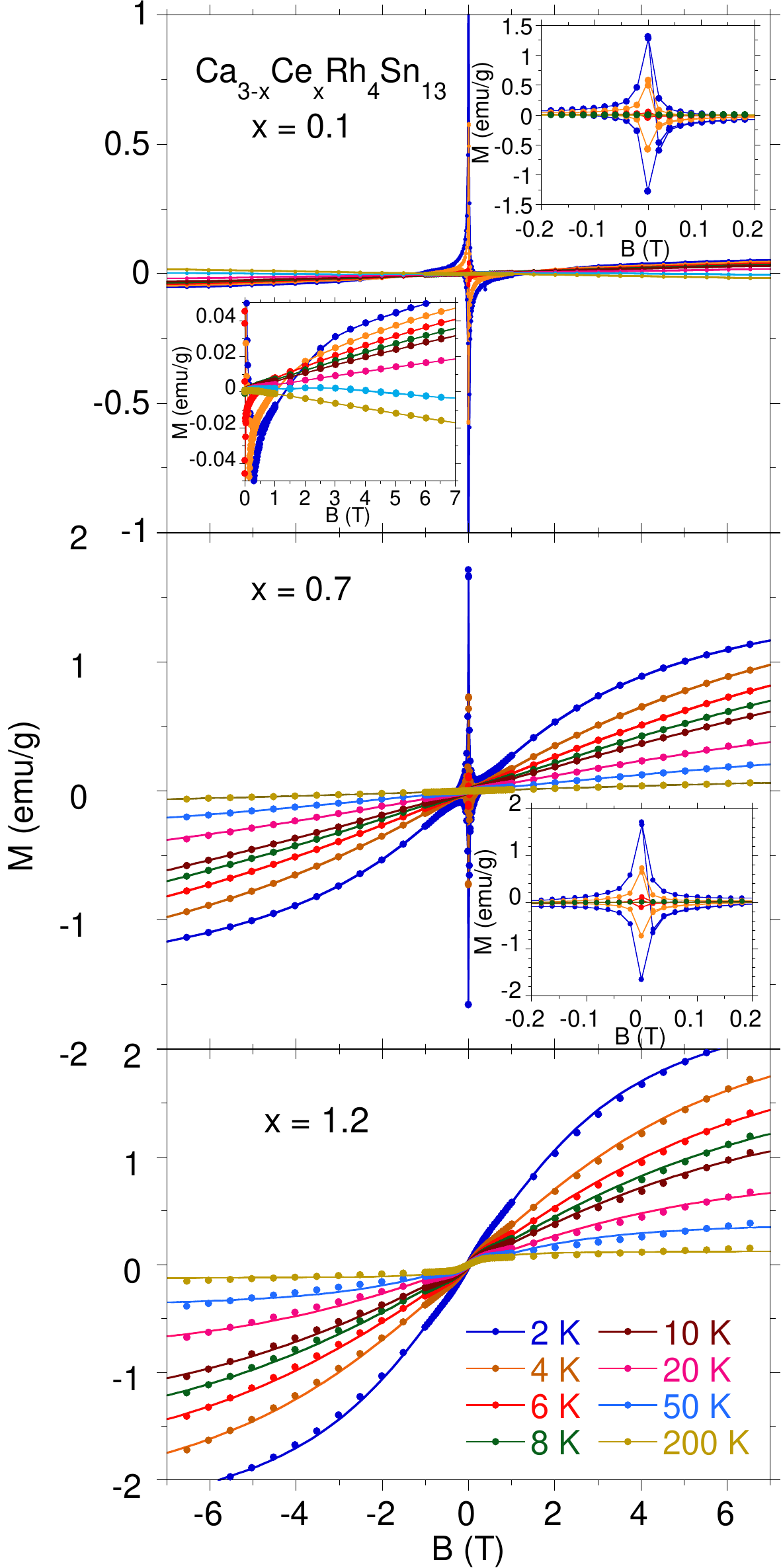}
\caption{\label{fig:Fig-M-vs-H_Ce_last}
Magnetization $M$ vs magnetic field $B$ for Ca$_{3-x}$Ce$_x$Rh$_{4}$Sn$_{13}$ at different temperatures. The insets exhibit hysteresis loops for superconducting state observed for the components $x<1.2$. A broad hysteresis loop suggests strongly inhomogeneous material. For $x\geq0.4$ the solid lines are fits of the Langevin function to the magnetization.
}
\end{figure}

\begin{figure}[h!]
\includegraphics[width=0.48\textwidth]{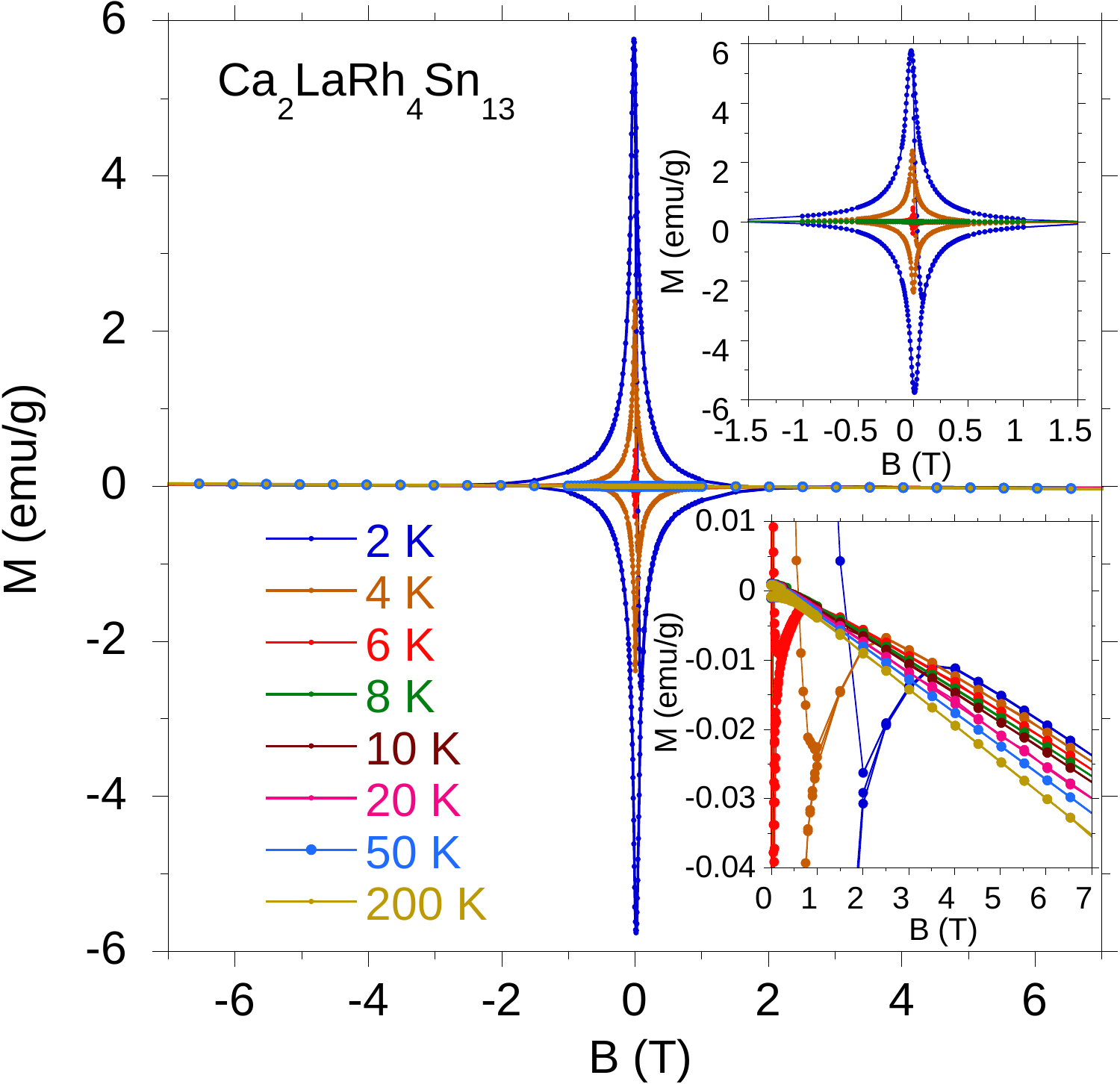}
\caption{\label{fig:Fig_M-H_La-1}
Magnetization $M$ vs magnetic field $B$ isotherms for Ca$_{2}$LaRh$_{4}$Sn$_{13}$ at different temperatures. Similar $M$ vs $B$ isotherms are observed for remaining Ca$_{3-x}$La$_x$Rh$_{4}$Sn$_{13}$ compounds. The insets exhibit hysteresis loops for superconducting state and the diamagnetic $M(B)$ behavior, respectively. A broad hysteresis loop suggests strongly inhomogeneous material. 
}
\end{figure}

Figures \ref{fig:Fig-M-vs-H_Ce_last} and \ref{fig:Fig_M-H_La-1} display the magnetization $M$ vs $B$ isotherms for Ca$_{3-x}$Ce$_x$Rh$_{4}$Sn$_{13}$ and Ca$_{2}$LaRh$_{4}$Sn$_{13}$ (a representative of the Ca$_{2}$LaRh$_{4}$Sn$_{13}$ family).
Ca$_{3-x}$La$_x$Rh$_{4}$Sn$_{13}$ alloys are diamagnetic in the wide temperature region with hysteresis loops representing the effect of vortex pinning, while the $M(B)$ isotherms for Ca$_{3-x}$Ce$_x$Rh$_{4}$Sn$_{13}$ are well approximated by Langevin function $L(\xi) = coth (\xi) - \frac{1}{\xi}$, where $\xi = \frac{\mu B}{k_BT}$ with total magnetic moment $\mu \approx 0.8-0.9$ $\mu_B$ per Ce atom obtained for the isotherms at 2 K. The hysteresis loop effect completely disappears for Ce content $x>1$.

\begin{figure}[h!]
\includegraphics[width=0.48\textwidth]{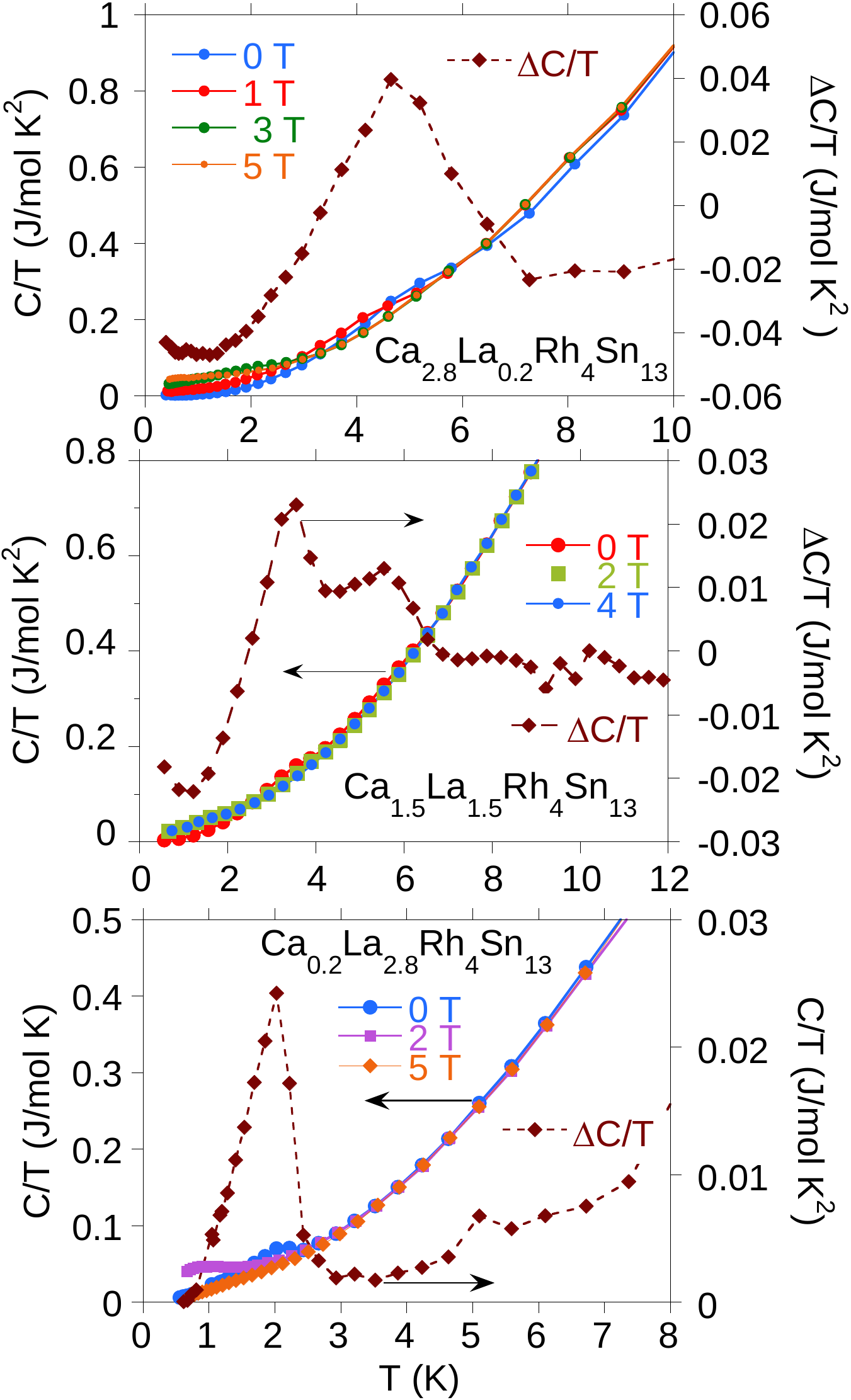}
\caption{\label{fig:Fig_C_DC-La-02-15-28_III}
Temperature dependence of specific heat, $C(T)/T$ and $\Delta C(T)/T$, for Ca$_{3-x}$La$_x$Rh$_{4}$Sn$_{13}$ in various magnetic fields. $\Delta C(T)/T$ is a difference of the $C(T)/T$ data obtained at the zero magnetic field and the field $B=5$ T. For Ca$_{2.8}$La$_{0.2}$Rh$_{4}$Sn$_{13}$, $\Delta C(T)/T$ has only one broad maximum, well approximated by the function  $f(\tilde{\Delta})$, while the components $x\geq 0.6$ clearly show in the $\Delta C(T)/T$ data two maxima due to presence of $T^{\star}_c$ and $T_c$ superconducting phases.
}
\end{figure}

Figure \ref{fig:Fig_C_DC-La-02-15-28_III} displays the specific heat data $C(T)/T$ and $\Delta C(T)/T$ for selected Ca$_{3-x}$La$_x$Rh$_{4}$Sn$_{13}$ compounds, the $\Delta C(T)/T$ is defined as a difference between the $C/T$ data measured at the zero magnetic field and at the field of 5 T. There is no sharp transition at $T_c$ in the specific heat data of the sample with $x=0.2$; instead, the specific heat displays a broad peak below $T_c$, which is strongly reduced by field. This $C(T)$ effect was attributed to the inhomogeneous {\it high temperature} superconducting  $T^{\star}_c$ phase due to 
atomic disorder \cite{Slebarski2014a,Slebarski2015}. 
It has been shown in Ref. \onlinecite{Andersen2006} that potential disorder smooth on a scale comparable to the coherence length leads to large modulation of the superconducting gap and large transition width \cite{comment3}. A simple Gaussian gap $\tilde{\Delta}$ distribution \cite{Slebarski2014a}
\begin{equation}
f(\tilde{\Delta})\propto \exp\left[-\frac{\left(\tilde{\Delta}-\tilde{\Delta}_0\right)^2}{2D}\right],
\end{equation}
where $\tilde{\Delta}_0$ and $D$ are treated as fitting parameters, well fits the $\Delta C(T)/T$ data for strongly disordered Ca$_{2.8}$La$_{0.2}$Rh$_{4}$Sn$_{13}$ alloy. The maximum of $f(\tilde{\Delta})$ distribution well agrees with the temperature of the $\chi^{''}$ maximum (3) in Fig. \ref{fig:Fig_CHI_ac_La-25}. The $C(T)/T$ behavior in this strongly disordered alloy is qualitatively different than that in reach in La Ca$_{3-x}$La$_x$Rh$_{4}$Sn$_{13}$ compounds with clear evidence for two superconducting phases: the {\it high temperature} inhomogeneous  superconducting $T^{\star}_c$ phase and the bulk superconducting state below $T_c$, where $T^{\star}_c > T_c$. Recently we noted  that  the $C(T)$ data for La$_{3}$Rh$_{4}$Sn$_{13}$ \cite{Slebarski2014a} and Ca$_{3}$Rh$_{4}$Sn$_{13}$ \cite{Slebarski2016} are well estimated by $C(T) \sim exp[-\frac{\Delta (0)}{k_BT}]$, which indicates that these parent compounds are $s$-wave superconductors and follows the behavior described by the BCS theory in the weak-coupling limit ($\Delta (0) $ is the energy gap of $T_c$ phase at zero temperature). 
The $C(T)$ data of cerium doped (Ca$_{3-x}$Ce$_{x}$Rh$_{4}$Sn$_{13}$) samples were recently  reported in Ref. \onlinecite{Slebarski2016}. 
It was documented \onlinecite{Slebarski2016} that the  broad maximum observed  at $T<T_c$ in the specific heat data of the samples $0<x<1.2$
represents an inhomogeneous superconducting phase in presence of spin-glass-like state, with evident contribution of the short-range magnetic correlations to entropy.
\begin{figure}[h!]
\includegraphics[width=0.48\textwidth]{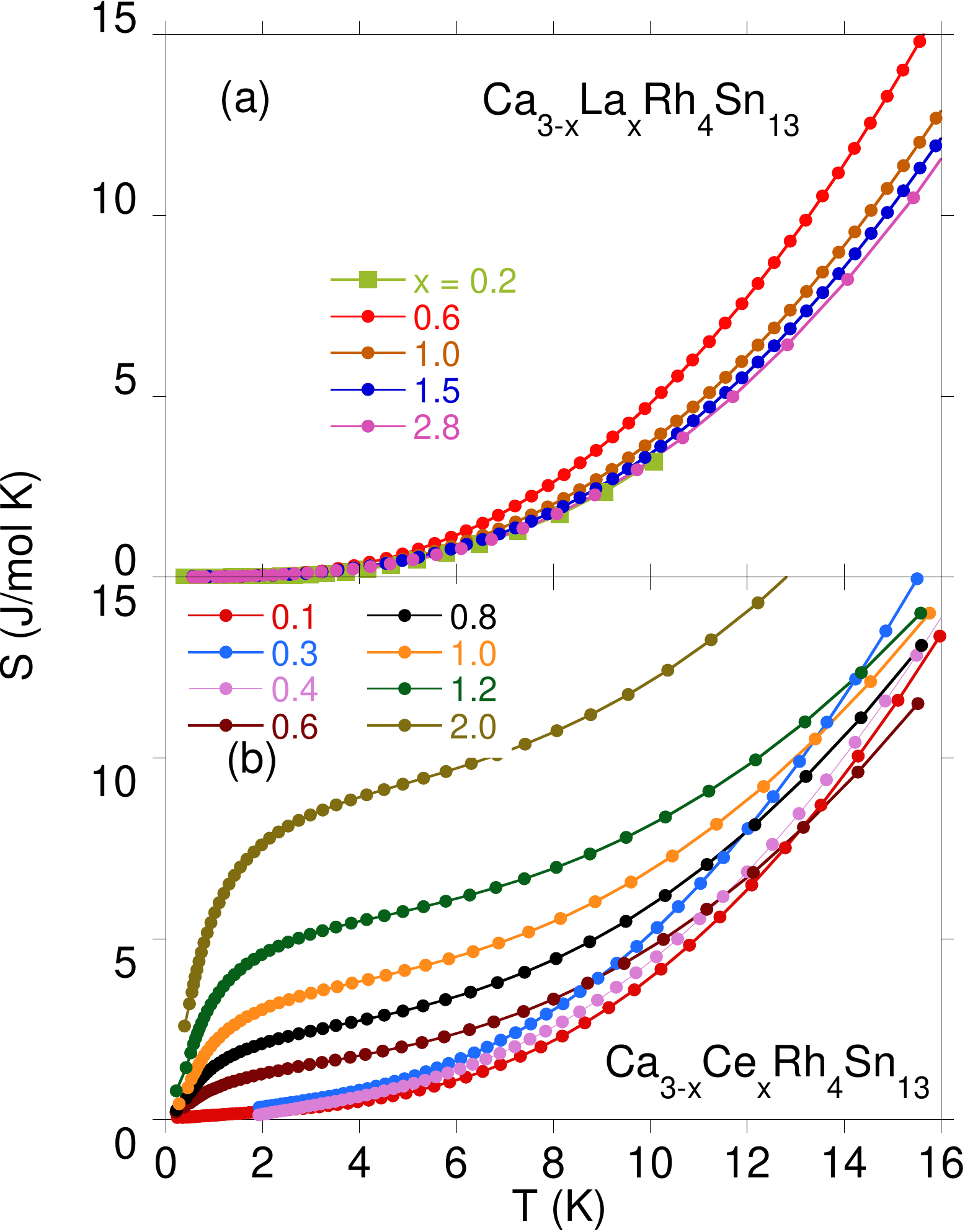}
\caption{\label{fig:S_sum}
The entropy $S$ of the samples Ca$_{3-x}$La$_{x}$Rh$_{4}$Sn$_{13}$ ($a$) and Ca$_{3-x}$Ce$_{x}$Rh$_{4}$Sn$_{13}$ ($b$) vs temperature at zero magnetic field $B$.
}
\end{figure}
Figure \ref{fig:S_sum} compares the entropy of Ca$_{3-x}$Ce$_{x}$Rh$_{4}$Sn$_{13}$ alloys with that of nonmagnetic La-doped alloys. 
Panel ($b$) displays the  magnetic entropy $S$ which  for $x\geq 0.6$ has a linear scaling with $x$, and is  
$Rln2$ per Ce atom at about $6-7$ K, indicating that the entropy represents the behavior of the ground state doublet. 
\begin{figure}[h!]
\includegraphics[width=0.48\textwidth]{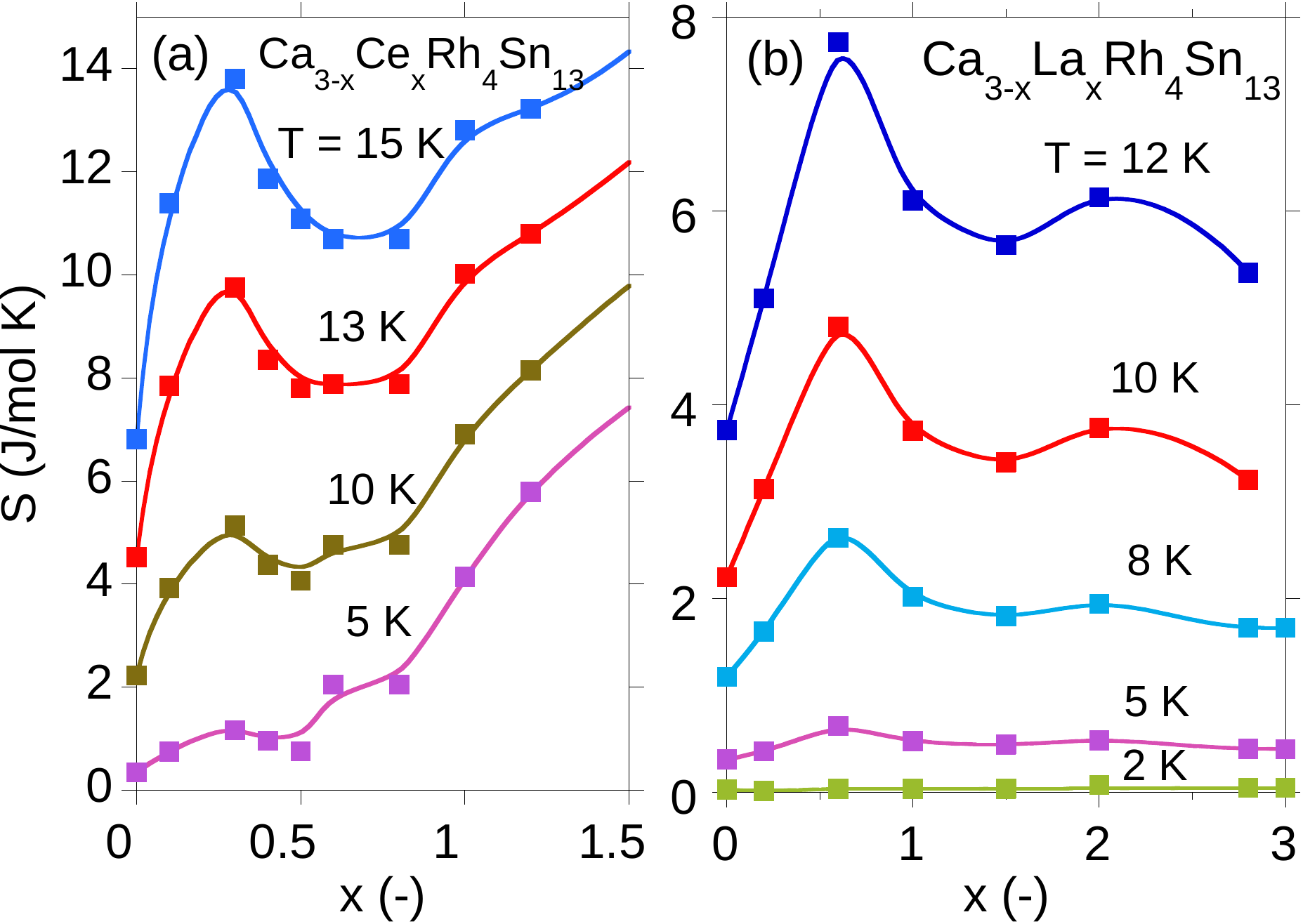}
\caption{\label{fig:Fig_S-vs-x_La-Ce-T-const}
Entropy $S$ isotherms as a function of doping $x$ for Ca$_{3-x}$Ce$_{x}$Rh$_{4}$Sn$_{13}$ ($a$) and 
Ca$_{3-x}$La$_{x}$Rh$_{4}$Sn$_{13}$ ($b$) at different temperatures.}
\end{figure}
Figure \ref{fig:Fig_S-vs-x_La-Ce-T-const} displays $S$ vs $x$ isotherms for Ce [panel ($a$)] and La [panel ($b$)] alloys. The $S(x)$ isotherms of Ca$_{3-x}$Ce$_{x}$Rh$_{4}$Sn$_{13}$ are plotted at different temperatures $T$ larger than the temperature of spin-glass-like ordering. In Fig. \ref{fig:Fig_S-vs-x_La-Ce-T-const}$a$ $S(x)_{T=const}$ increases  with  Ce substitution, which can  easily be explained as a result of  the $x$ dependent paramagnetic spin-disorder effect (c.f. Fig. \ref{fig:Fig-M-vs-H_Ce_last}). This behavior was not observed  for the {\it nonmagnetic} La-doped series of alloys, as is shown in  Fig.  \ref{fig:Fig_S-vs-x_La-Ce-T-const}$b$. 
In addition, for $x = 0.3$ and 0.8 when Ca$_{3}$Rh$_{4}$Sn$_{13}$ is substituted by Ce, and for $x = 0.6$ and $\sim 2$  for La doping, the isotherms $S(x)_{T=const}$ show a clear maxima that correspond 
to the largest separation between the  superconducting phases 
$T^{\star}_c $ and $T_c $  due to atomic disorder, as is displayed in $T-x$ diagram
(will be discussed in  section  A.3.).

\subsubsection{$T_c - x$ phase diagrams for Ca$_{3}$Rh$_{4}$Sn$_{13}$ doped with La and Ce}

\begin{figure}[h!]
\includegraphics[width=0.48\textwidth]{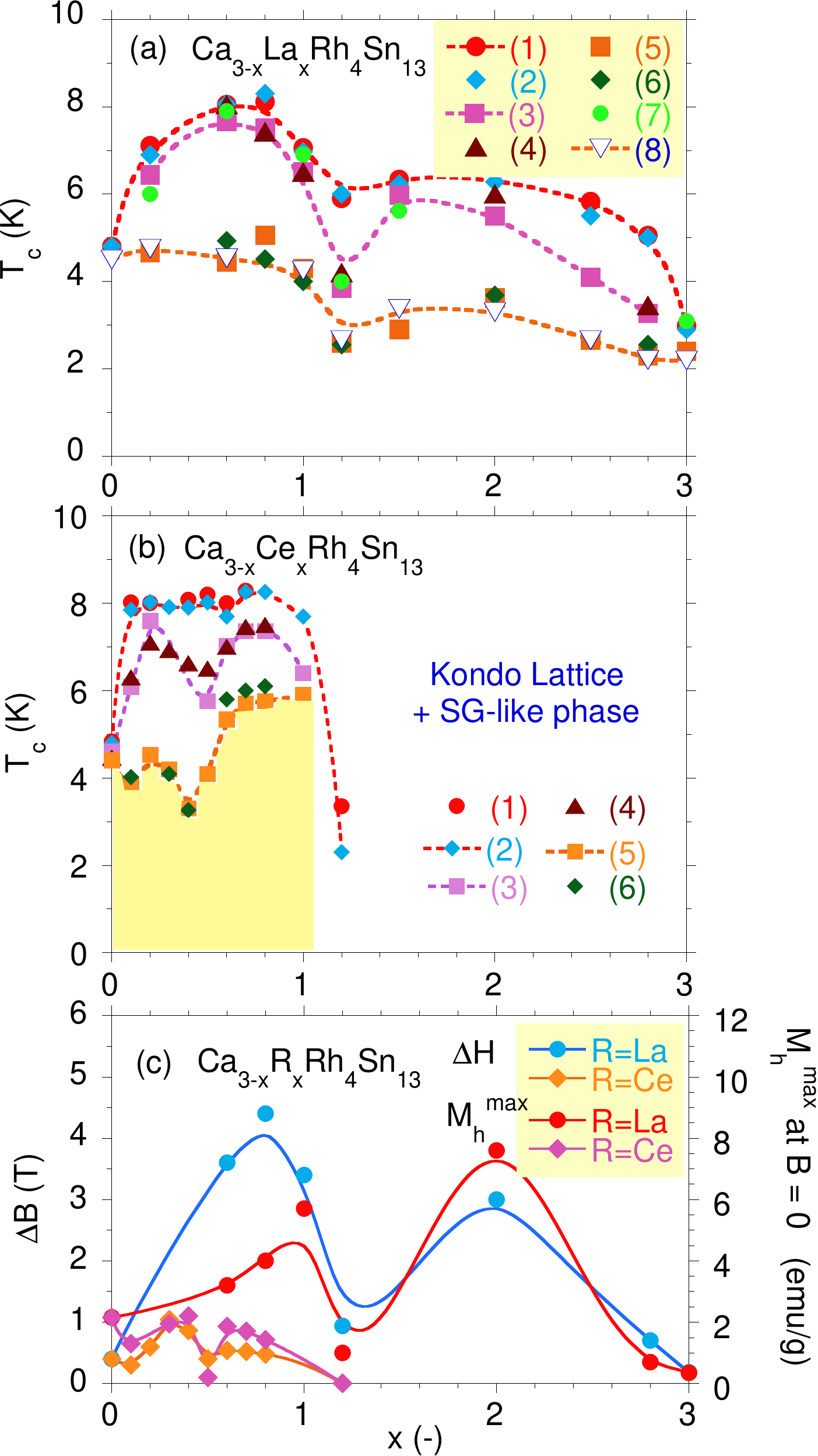}
\caption{\label{fig:Fig_DIAGRAM_T-x-sum}
$T-x$ $T^{\star}_c$ and $T_c$ phase diagram of Ca$_{3-x}$La$_x$Rh$_4$Sn$_{13}$ ($a$) and Ca$_{3-x}$Ce$_x$Rh$_4$Sn$_{13}$ ($b$) compounds from electrical resistivity (1), ac susceptibility [points (2), (3), and (5)], dc magnetic susceptibility [points (4) and (6)], and specific heat [points (7) and (8)] measurements (details in the text).
The dotted red curve represents  the critical temperatures at which
the $T^{\star}_c $ inhomogeneous superconducting phase begins to be formed. The purple dotted line represents the $T^{\star}_c $ inhomogeneous phases with superconducting gaps $\tilde{\Delta_0}$ corresponding to the maximum of Gaussian gap distribution $f(\tilde{\Delta})$.
The critical temperatures $T_c$ of the bulk superconducting phases are represented by the dotted orange line. Panel ($c$) shows details of the hysteresis loop effect in the superconducting regime; $\mid M_h^{max}\mid$ is a maximum value of the magnetization $M$, and $\Delta B$ is the maximum field where the hysteresis loop is observed (c.f. Figs. \ref{fig:Fig_M-H_La-1} and \ref{fig:Fig-M-vs-H_Ce_last}). 
} 
\end{figure}

In summary, we present in Fig. \ref{fig:Fig_DIAGRAM_T-x-sum} a $T-x$ diagram of the superconducting $T^{\star}_c$ and $T_c$ phases for Ca$_{3}$Rh$_{4}$Sn$_{13}$ doped with La (panel $a$) and Ce (panel $b$). 
The comprehensive magnetic, electrical resistivity, and specific heat study suggest coexistence of  short-range magnetic correlations with superconductivity in the Ce-doping regime $x<1.2$ this is, however,  not a case of Ca$_{3}$Rh$_{4}$Sn$_{13}$ doped with La. 
In Fig. \ref{fig:Fig_DIAGRAM_T-x-sum}  points (1) are obtained  at 50\% of the normal state resistivity value (c.f. Fig. \ref{fig:Fig_R-H_sum_La-02-15-28}). The temperatures of the respective maxima in $d\chi^{''}/dT$ shown in Fig. \ref{fig:Fig_CHI_ac_Ce-sum}  are presented as the points (2), (3) and (5). The $\chi$ versus $T$ dc magnetic susceptibility data obtained at 500 Oe in zero field (ZFC) and field cooling (FC) modes reveals the onset of diamagnetism and thermal hysteresis associated with the superconducting state below $T^{\star}_c $ [points (4)]. Points (6) represent temperature of the maximum in $d\chi_{dc}/dT$, which is $T$ of about  $1/2$ of the diamagnetic dc susceptibility drop.
Finally, points (7) and (8) represent $T^{\star}_c $ and $T_c $ obtained from specific heat $C(T)/T$ data, respectively (c.f. Fig.  \ref{fig:Fig_C_DC-La-02-15-28_III}). The $T-x$ diagram clearly indicates the presence of two separate superconducting phases, $T^{\star}_c $ and $T_c $. An increase of atomic disorder enhances the separation between them and the largest one is obtained for the La substituted samples with $x\sim 0.6$ and for the Ce substituted  Ca$_{3}$Rh$_{4}$Sn$_{13}$ alloys with $x\sim 0.3$. We also documented that spin-glass-like magnetic correlations increases  $T^{\star}_c $ and $T_c $ for the Ca$_{3-x}$Ce$_x$Rh$_{4}$Sn$_{13}$ compounds with respect to Ca$_{3-x}$La$_x$Rh$_{4}$Sn$_{13}$. This observation is interesting, and previously was motivated by theory \cite{Larkin1971,Galitski2002}. 
It was theoretically documented that the superconducting transition temperature is higher in the presence of the spin-spin interactions of the magnetic impurities, which form a spin-glass state.

The $T-x$ diagram also shows the minimum for $T_c$ versus $x$ dependence at $x_{min}\sim 1.2$ for Ca$_{3-x}$La$_x$Rh$_{4}$Sn$_{13}$, and at $x\sim 0.4$ for Ca$_{3-x}$Ce$_x$Rh$_{4}$Sn$_{13}$, however,  in the both cases $x_{min}$ is about 40\% of the whole superconducting $x$-region.  

\subsection{Electronic structure of Ca$_{3}$Rh$_{4}$Sn$_{13}$ doped with La and Ce; experiment and calculations}

\begin{figure}[h!]
\includegraphics[width=0.48\textwidth]{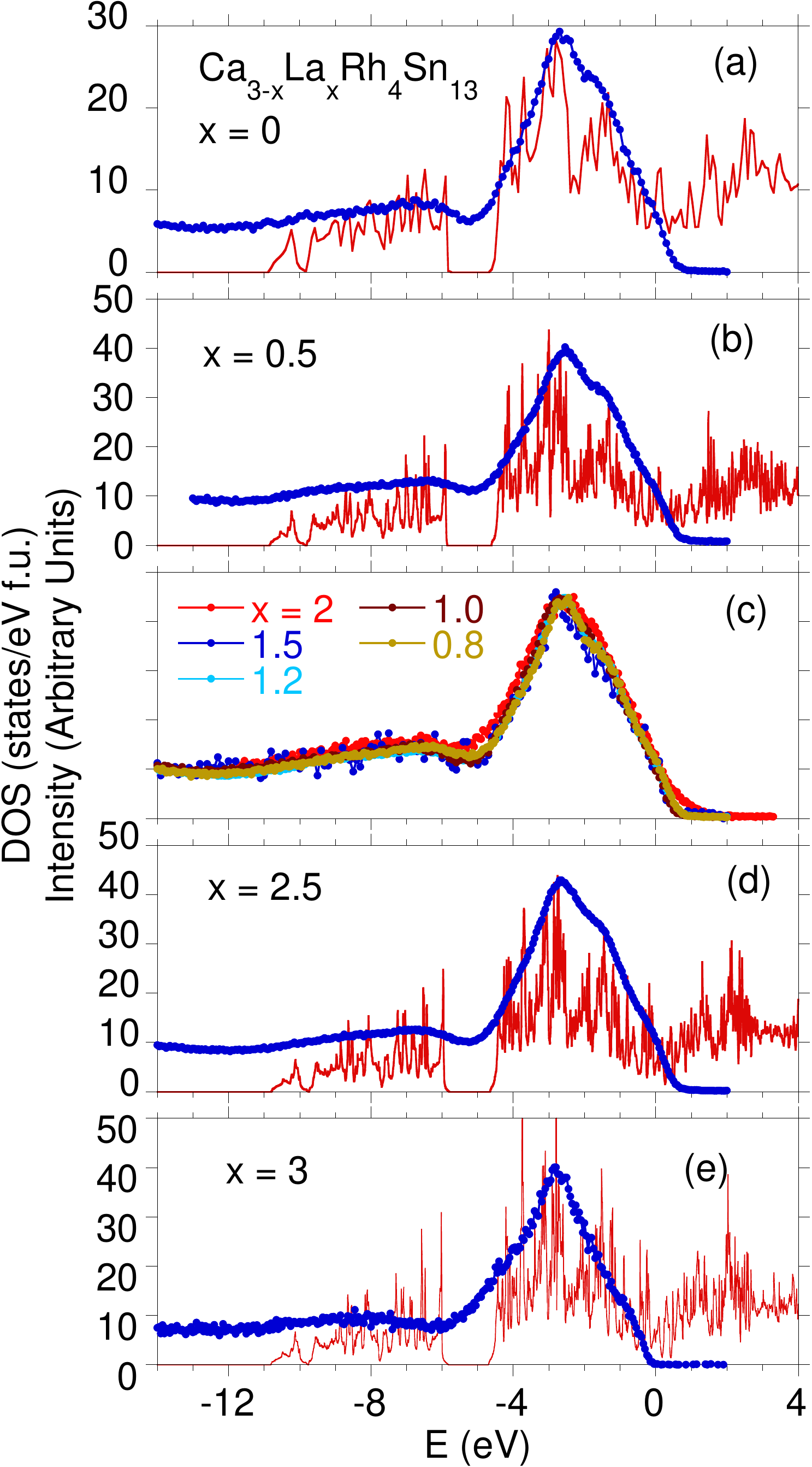}
\caption{\label{fig:Fig_DOS-XPS_sumLa}
Valence band XPS spectra for Ca$_{3-x}$La$_x$Rh$_4$Sn$_{13}$ compared with the calculated 
total density of states  within the LSDA approximation for the components $x=0$, 0.5, 2.5, and 3. The figure also shows the VB XPS spectra for intermediate components of the series [in panel ($c$)]. 
} 
\end{figure}

In order to explain the alloying effect on the band structure of Ca$_{3}$Rh$_{4}$Sn$_{13}$ superconductor, we investigated  valence-band (VB) XPS spectra of the Ca$_{3-x}$La$_x$Rh$_4$Sn$_{13}$ and Ca$_{3-x}$Ce$_x$Rh$_4$Sn$_{13}$ samples. We also analyze the Sn $4d$ XPS core-level spectra to demonstrate the nature of the covalent bonding between Sn1 and Sn2 atoms in Sn1{Sn2}$_{12}$ cages.
Figure \ref{fig:Fig_DOS-XPS_sumLa} shows the VB XPS spectra for the series of Ca$_{3-x}$La$_x$Rh$_4$Sn$_{13}$ alloys. The XPS bands for the components $x=0$, 0.5, 2.5, and 3 are compared with calculated total DOSs. 
Panel ($c$)  displays also the VB XPS spectra obtained for intermediate components $0.5<x<2.5$. The valence band XPS spectra of Ca$_{3-x}$Ce$_x$Rh$_4$Sn$_{13}$ alloys were recently presented in Ref. \onlinecite{Slebarski2016}, therefore are not shown here.
All spectra measured for both series are very similar and dominated by Rh $4d$ electron states, which are located in the XPS bands between the Fermi energy $\epsilon_F$ and $\sim 4$ eV, 
and by Sn $5s$ states with the broad maximum centered at $\sim 7$ eV. With increasing concentration of La or Ce, the shape of the VB XPS spectra are almost the same, excluding the narrow energy range near $\epsilon_F$, which strongly relates to the electric transport properties of these alloys. Recently, we have demonstrated  that, for  metallic state of the Ca$_{3-x}$Ce$_x$Rh$_4$Sn$_{13}$ alloys, the subtle change of DOS at $\epsilon_F$ correlates well with the observed resistivity behavior $\frac{d\rho}{dP}$ giving  $\frac{dN(\epsilon_F)}{dP} \propto \frac{d\rho}{dP}$~\cite{Slebarski2016}.  A similar effect was observed for Ca$_{3}$Rh$_4$Sn$_{13}$ with Ca atoms partially replaced by La. 
\begin{figure}[h!]
\includegraphics[width=0.48\textwidth]{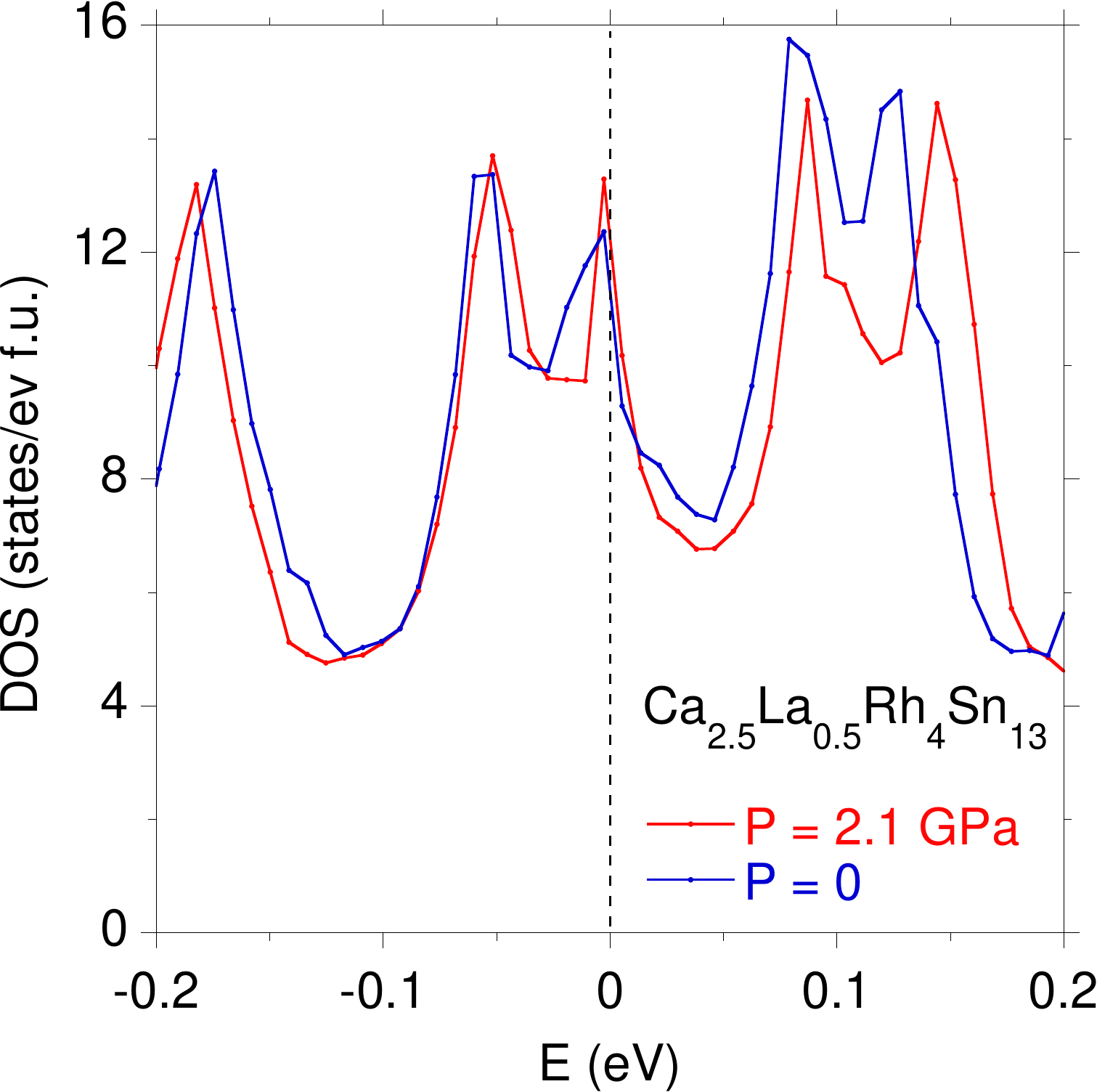}
\caption{\label{fig:Fig_DOS-La_vs_P}
The total DOS near the Fermi energy calculated for Ca$_{2.5}$La$_{0.5}$Rh$_4$Sn$_{13}$ at the pressure 0 and 2.1 GPa. 
} 
\end{figure}
Figure \ref{fig:Fig_DOS-La_vs_P} demonstrates the change of calculated DOS at $P = 0$ and 2.1 GPa for Ca$_{2.5}$La$_{0.5}$Rh$_4$Sn$_{13}$. Calculations documented the increase of the total DOS at $\epsilon_F$ with $P$, giving $\frac{dN(\epsilon_F)}{dP}\cong -0.5$ eV$^{-1}$ GPa$^{-1}$, which correlates well with the observed negative $\frac{d\rho}{dP}$ in normal metallic state  at $T=8$ K, as shown in Fig. \ref{fig:Fig_R-vs-P_La-02_last}. This simple explanation assumes the relation $\rho \sim 1/n$ between the resistivity and the number of carriers $n$, that naively reflects the DOS at $\epsilon_F$. Figure \ref{fig:Fig_DOS-La_vs_P} also shows the energy shift of the DOS maxima with $P$ in the vicinity of $\epsilon_F$ to higher binding energy $\mid E \mid$ in respect to $\epsilon_F$. 
This response for heavy Fermi metals to the applied pressure is characteristic of electron-type conductivity at high pressure \cite{Hai1999}.

Figure \ref{fig:Fig_R-vs-P_La-25_last} exhibits at $T>T_c$ 
positive effect of $\frac{d\rho}{dP}$  for  Ca$_{0.5}$La$_{2.5}$Rh$_4$Sn$_{13}$ samples rich in La. This $\frac{d\rho}{dP}>0$ behavior, different  with that observed for La-diluted Ca$_{3}$Rh$_4$Sn$_{13}$ alloys, was recently discussed as a result of interband distance (pseudogap) located at the Fermi level, which increases with pressure (for details see Ref. \onlinecite{Slebarski2015b}), and leads to increase in $\rho$ under pressure.
\begin{figure}[h!]
\includegraphics[width=0.48\textwidth]{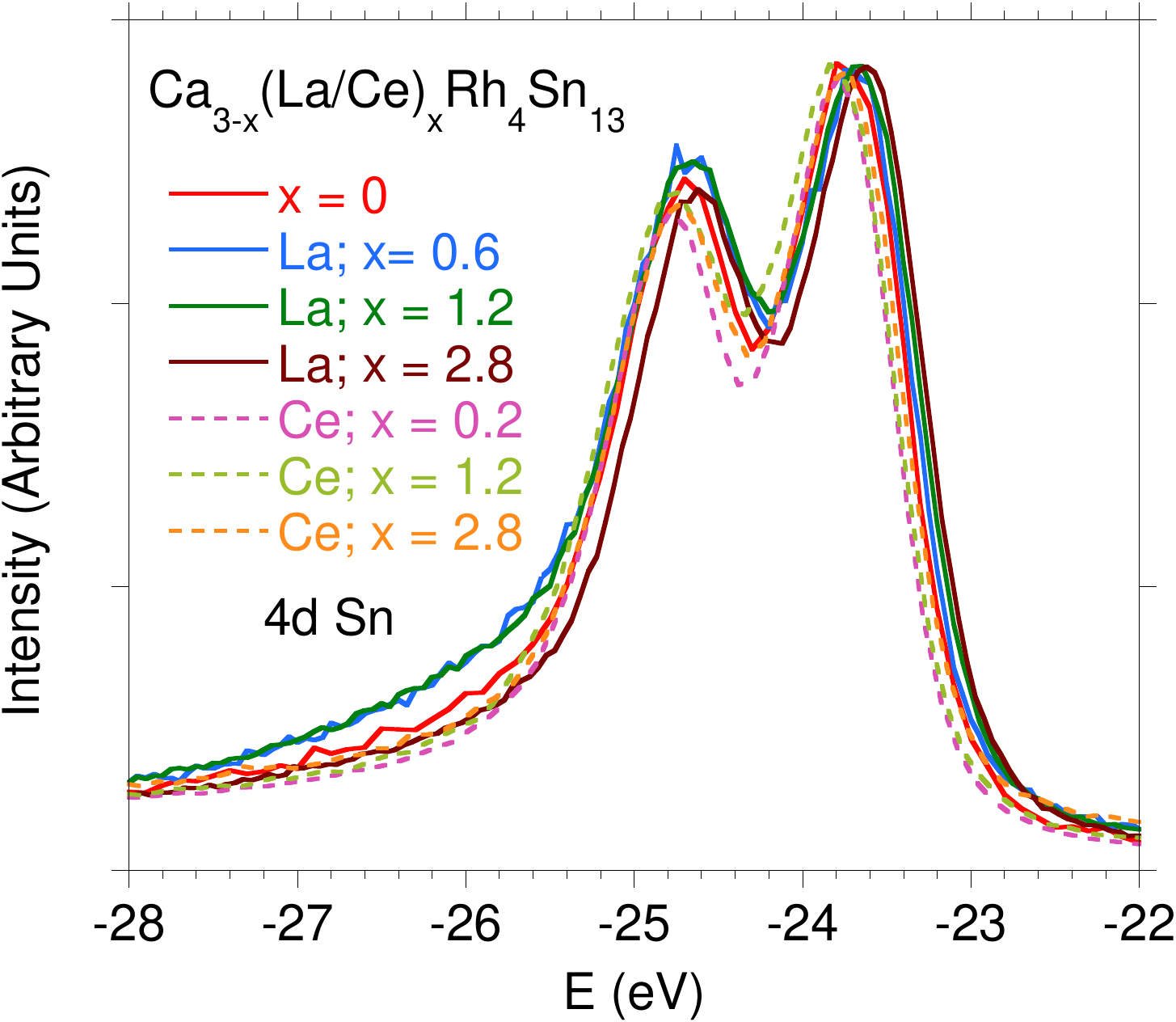}
\caption{\label{fig:Fig_Sn4d_La-Ce}
The Sn $4d$ XPS spectra for the series of Ca$_{3-x}$La$_{x}$Rh$_4$Sn$_{13}$ and Ca$_{3-x}$Ce$_{x}$Rh$_4$Sn$_{13}$ alloys. 
} 
\end{figure}
To signal the effect of covalent bonding between Sn1 and Sn2 atoms in the Sn1{Sn2}$_
{12}$ cage, we present Sn $4d$ XPS spectra in Fig. \ref{fig:Fig_Sn4d_La-Ce}. Recently \cite{Slebarski2013} we suggested that the asymmetry of the Sn $4d$ XPS spectrum may be associated with the change of local symmetry of the Sn2$_{12}$ cage due to strong  covalent bonding, manifested by lower than cubic local crystal electric field visible in susceptibility and specific heat data, and by structural distortion. The Sn $4d$ XPS spectra consist of two peaks at 23.8 eV and 24.7 eV binding energy with a spin-orbit splitting of $\sim 1 $ eV, and a broad feature with maximum at about 26 eV. The complex electronic structure of the Sn $4d$ XPS lines confirms the charge accumulation between Sn2 and Sn1, obtained from the calculated difference charge density for La$_3$Rh$_4$Sn$_{13}$ \cite{Gamza2008}. 

\subsubsection{Bonding properties investigated by electron localization function}

\begin{figure}[h!]
\includegraphics[width=0.48\textwidth]{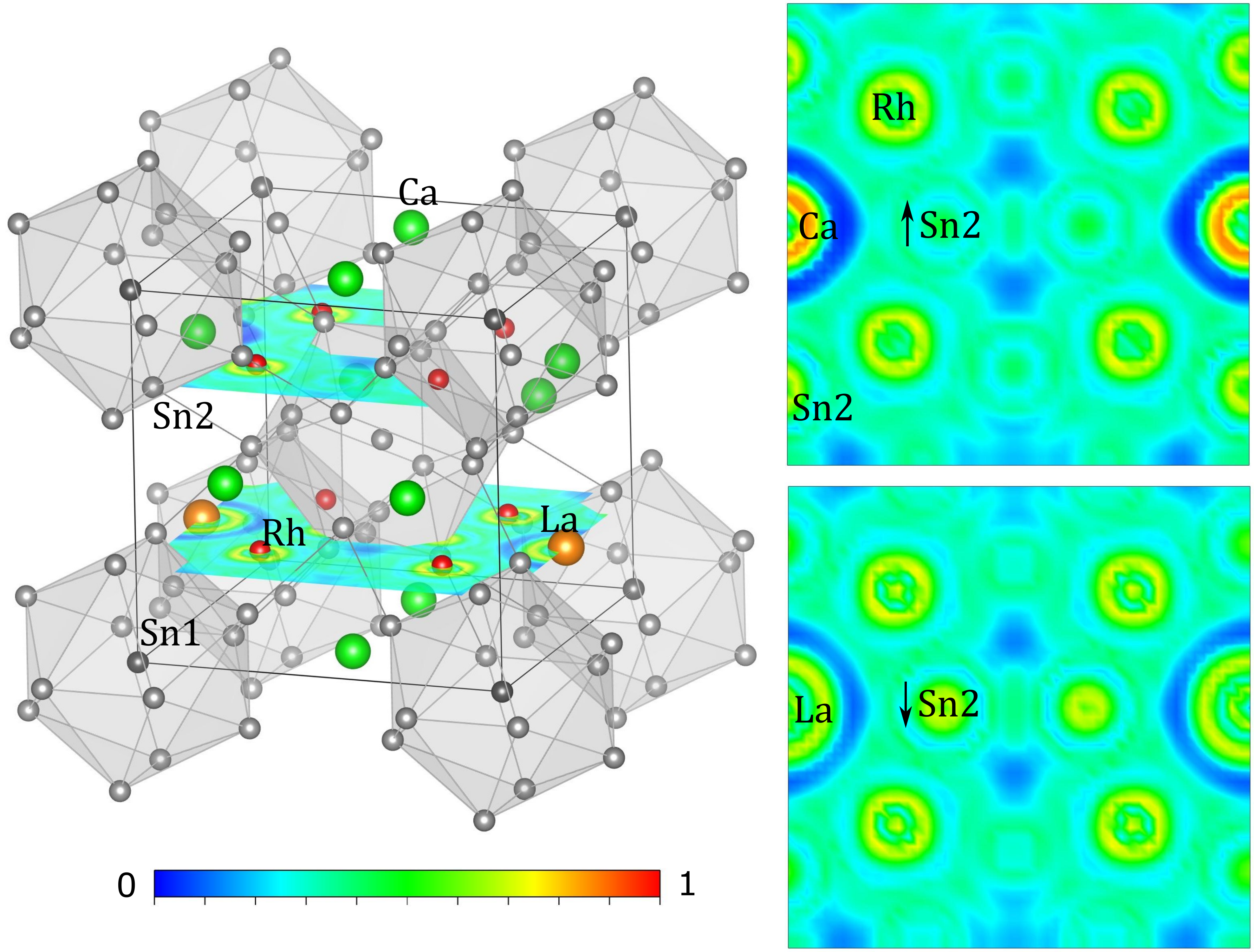}
\caption{\label{fig:5Ca1La_2planes_v2}
The ELF distribution for Ca$_{2.5}$La$_{0.5}$Rh$_4$Sn$_{13}$ in the plane $z=\frac{3}{4}$ shown in the upper panel and  in the plane  $z=\frac{1}{4}$ (in lower panel). An arrow oriented up/down indicates Sn2 atom located above or below the plane.
}
\end{figure}

\begin{figure}[h!]
\includegraphics[width=0.48\textwidth]{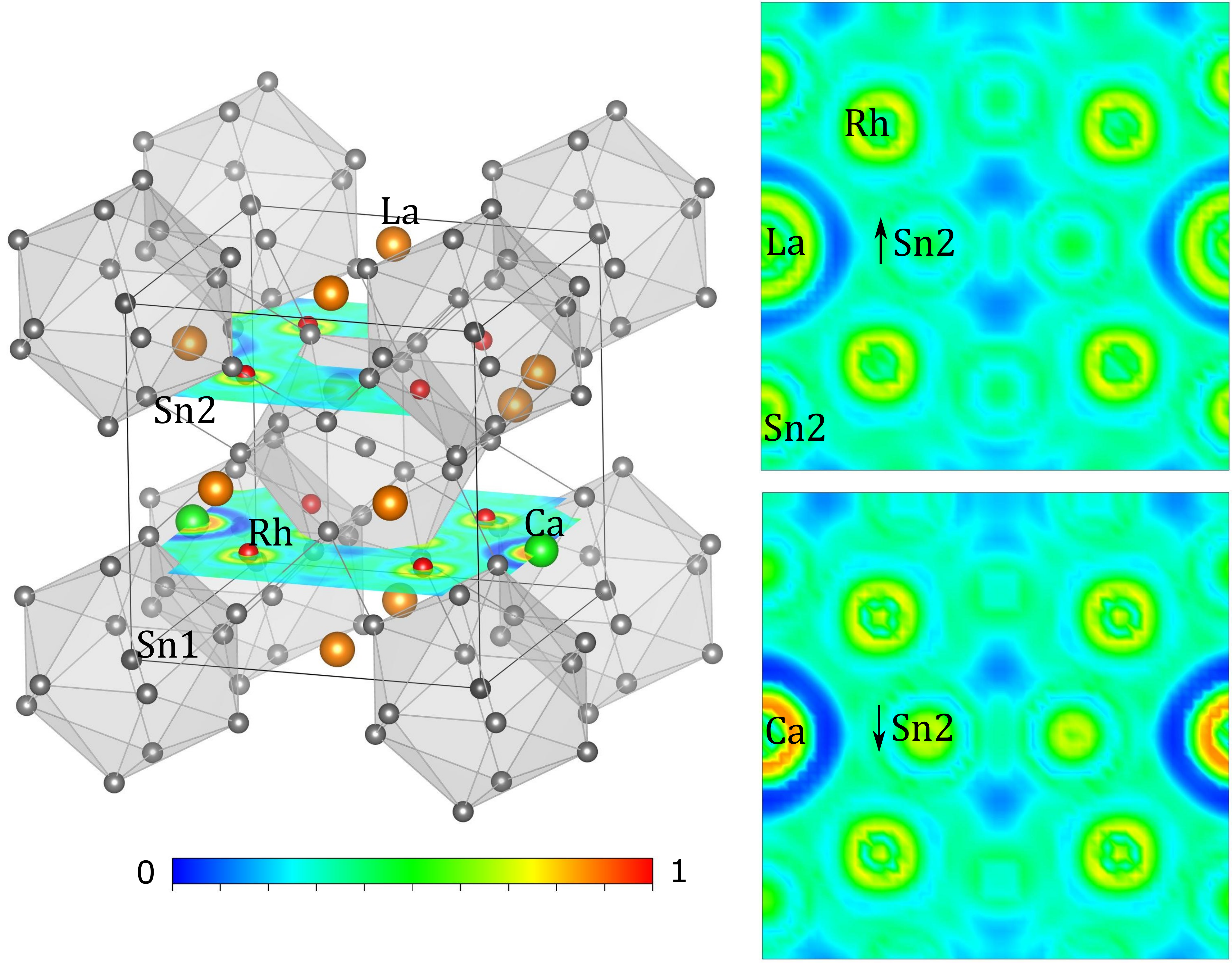}
\caption{\label{fig:5La1Ca_2planes_v2}
The ELF distribution for Ca$_{0.5}$La$_{2.5}$Rh$_4$Sn$_{13}$ in the plane $z=\frac{3}{4}$ shown in the upper panel and  in the plane $z=\frac{1}{4}$ (in lower panel). An arrow oriented up/down indicates Sn2 atom located above or below the plane.
}
\end{figure}

\begin{figure}[h!]
\includegraphics[width=0.48\textwidth]{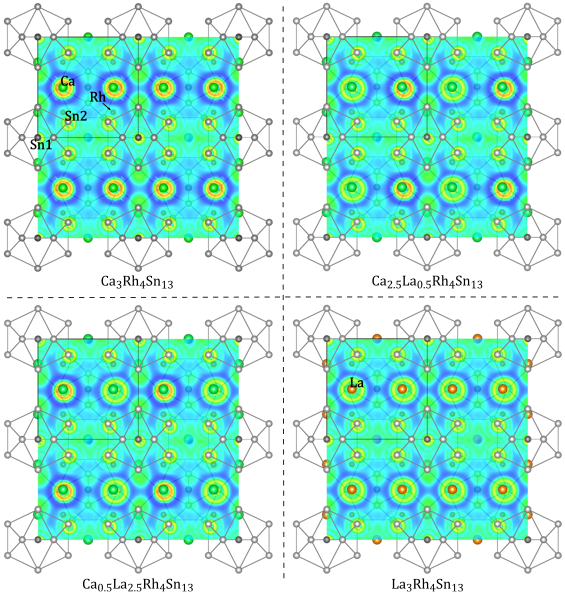}
\caption{\label{fig:Fig_Ca_La_comparison_v2}
The ELF distribution in the plane $z=1$ shown for series of compounds.
}
\end{figure}

For the skutterudite-related Ce$_3M_4$Sn$_{13}$ and La$_3M_4$Sn$_{13}$ compounds, where $M$ is a $d$-electron metal, the charge density analysis revealed a strong charge accumulation between Sn1 atom and  Sn2 atoms of the {Sn2}$_{12}$ cage, and between metal $M$ or rare earth element and Sn2 atoms, which implies a strong covalent bonding interaction and leads to a subtle structural transition \cite{Slebarski2014,Slebarski2015a,Slebarski2015b}.
The structural deformation observed in this class of materials at $T^{\star} \sim 170$ K, is usually accompanied by  formation of a charge density wave (CDW) phase transition, and under external pressure $T^{\star}\longrightarrow 0$ defines a novel structural quantum critical point \cite{Klintberg2012}.  Based on the  present structural data, which documented the second order structural phase transition at much higher temperature of about 310 K, we revise the previous understanding of the nature of $T^{\star}\sim 170$ K (section III.C.).
The structural deformation, however, has  not been documented for Ca$_{3}$Rh$_4$Sn$_{13}$. 
To determine the subtle bonding properties of 
the charge distribution in La substituted Ca$_3$Rh$_4$Sn$_{13}$  a full-potential chemical-bonding analysis via calculation of the electron localization function (ELF) 
within the density functional theory \cite{Hohenberg1964} [similar band structure ELF calculations were presented for cerium doped  (Ca$_{1-x}$Ce${x}$Rh$_4$Sn$_{13}$) compounds recently in Ref. \onlinecite{Slebarski2016}]. 
Figures \ref{fig:5Ca1La_2planes_v2} and  \ref{fig:5La1Ca_2planes_v2} exhibit the ELF distribution in planes $z=\frac{3}{4}$ and $z=\frac{1}{4}$ for Ca$_{2.5}$La$_{0.5}$Rh$_4$Sn$_{13}$ and Ca$_{0.5}$La$_{2.5}$Rh$_4$Sn$_{13}$, respectively, while Fig. \ref{fig:Fig_Ca_La_comparison_v2} compares the ELF isosurfaces  $z=1$ sections for parent and doped compounds.
The ELF maxima are essentially located on the atoms in the plane, the charge density analysis also reveals the covalent bondings between the nearest-bounding atoms: Sn1-Sn2,  Rh-Sn2 and La-Sn2  due to charge accumulation between them. These  bondings are the strongest between Rh-Sn2 and Sn1 and Sn2, which can be a reason  of structural distortion. XRD analysis confirms the charge modulation on the Rh atoms. As a result of this charge modulation, some of the Sn2 atoms are located closer to Rh than the remaining one, which leads to distortion of the Sn2 cages. Moreover, when Ca in Ca$_3$Rh$_4$Sn$_{13}$ is fractionally replaced by larger La or Ce atom, the Sn$_{12}$ cage can easily be deformed due to strong chemical stresses, as is shown in Fig. \ref{fig:Fig_Ca_La_comparison_v2} (c.f. Fig. \ref{fig:Fig1_a-x-La-Ce}).
The superlattice transition that is connected with Sn2 cage instability is, however,  not observed in Ca$_3$Rh$_4$Sn$_{13}$.

\subsection{New insights into structural properties of Ca$_{3-x}$La$_{x}$Rh$_4$Sn$_{13}$ and Ca$_{3-x}$Ce$_{x}$Rh$_4$Sn$_{13}$ at high temperatures }

There has been only one type of structure related transition reported in the 3:4:13 system, related to the appearance of the modulation with a single arm \cite{Oswald2017} or a whole k-star\cite{Mazzone2015,Otomo2013} of the propagation vector $k =$ ($\frac{1}{2}$, $\frac{1}{2}$, 0). Transition temperatures associated with an appearance of the superstructure, commonly called $T^{\star}$, were  usually lower than 150~K and could be suppressed to QCP by doping\cite{Goh2015}.\\ The first round of temperature dependent XRD aimed for screening any similar effects in both series. As a starting crystallographic model, a simple cubic cell (SG {\it Pm}$\bar{3}${\it n} No. 223, without the superstructure) was selected and atom occupancies were constrained to the nominal values. Atom positions were chosen as La/Ca/Ce (6d) ($\frac{1}{4}$, $\frac{1}{2}$, 0), Rh (8e) ($\frac{1}{4}$, $\frac{1}{4}$, $\frac{1}{4}$), Sn1 (2a) (0, 0, 0), Sn2 (24k) (0, {\it y$_
{Sn2}$}, {\it z$_{Sn2}$}). All sites were refined with individual isotropic atomic displacement parameters (ADP) B$_{iso}$ \cite{CommentPZ1}.

\subsubsection{Ca-Ce series}
\begin{figure}[h!]
\includegraphics[width=0.48\textwidth]{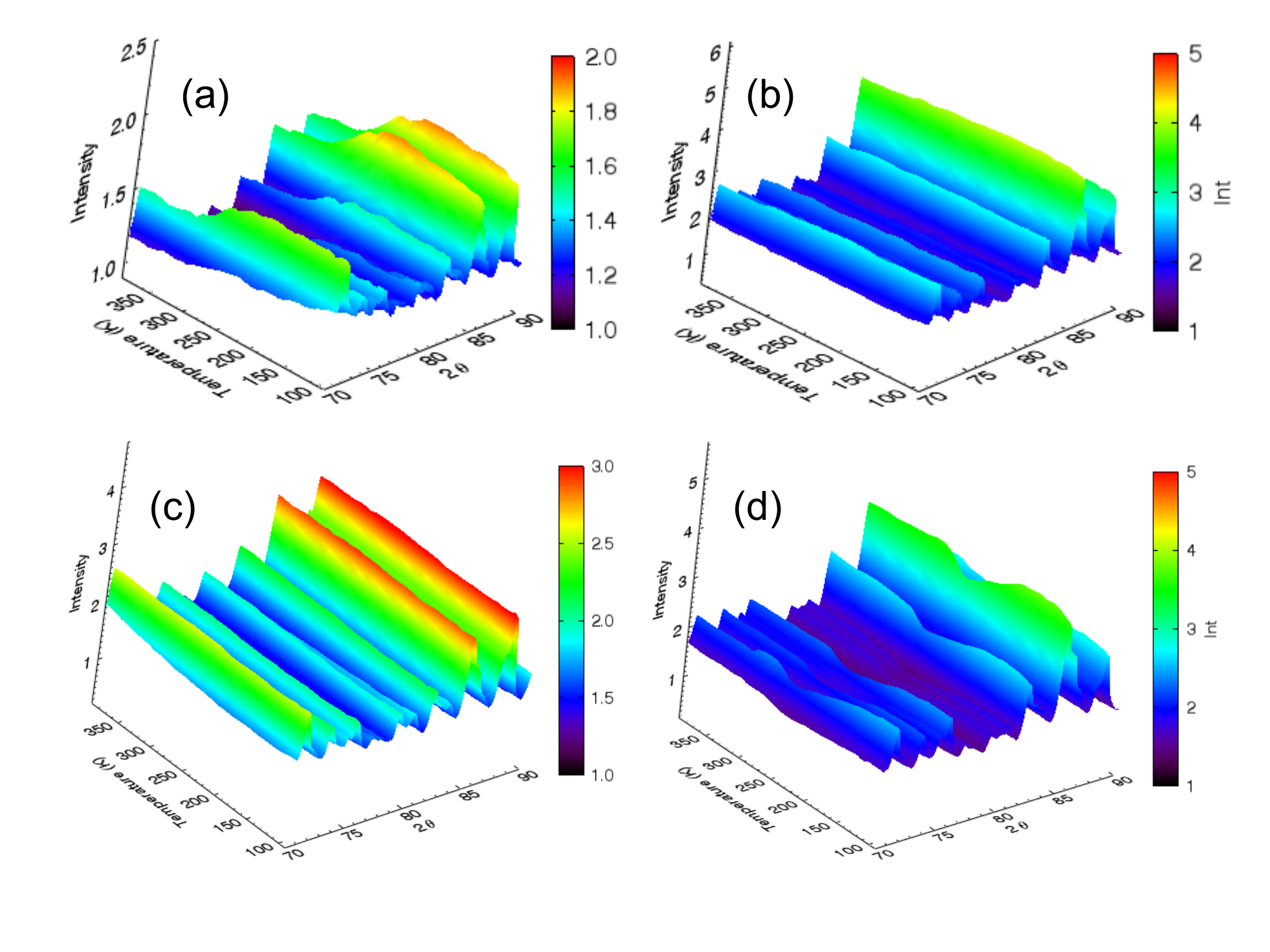}
\caption{\label{fig:Fig_Waterfalls}
Temperature dependence of middle sections of diffraction patterns for (a) Ca$_{1}$Ce$_{2}$Rh$_4$Sn$_{13}$, (b) Ca$_{2}$Ce$_{1}$Rh$_4$Sn$_{13}$, (c) Ca$_{1}$La$_{2}$Rh$_4$Sn$_{13}$ and (d) Ca$_{2.8}$La$_{0.2}$Rh$_4$Sn$_{13}$.
}
\end{figure}

In the Ca$_{3-x}$Ce$_{x}$Rh$_4$Sn$_{13}$ system we have studied two compositions with $x=1$ and $x=2$. Figure \ref{fig:Fig_Waterfalls}  presents middle sections of the diffraction patterns, which clearly shows presence of a possible transition for the Ce rich side (a) $x=2$ and uniform patterns in the other case (b) $x=1$. The experimental setup did not allow to observe any superstructure reflections.
In all cases, the variations in the intensities of reflections were accompanied by a prominent changes in ADPs (Fig. \ref{fig:Fig_Biso} (a) and (b)). Other parameters varied smoothly with the temperature (see Supplemental Information).

\begin{figure}[h!]
\includegraphics[width=0.48\textwidth]{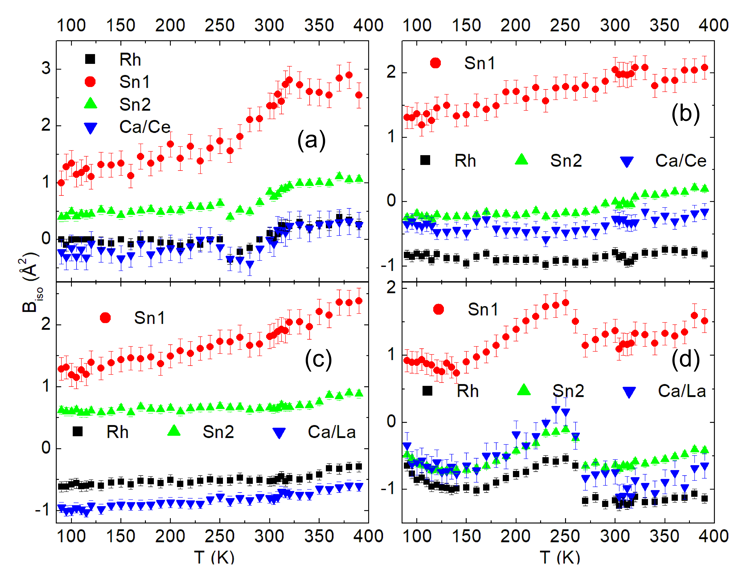}
\caption{\label{fig:Fig_Biso}
Temperature dependence of isotropic atomic displacement parameters (ADPs) for (a) Ca$_{1}$Ce$_{2}$Rh$_4$Sn$_{13}$, (b) Ca$_{2}$Ce$_{1}$Rh$_4$Sn$_{13}$, (c) Ca$_{1}$La$_{2}$Rh$_4$Sn$_{13}$ and (d) Ca$_{2.8}$La$_{0.2}$Rh$_4$Sn$_{13}$.
}
\end{figure}
An evidence of a possible transition is clearly seen for CaCe$_2$Rh$_4$Sn$_{13}$, where a gradual loss on intensity occurs in the 250 K to 300 K range (Fig. \ref{fig:Fig_Waterfalls} (a)). At the same time, the intensities for Ca$_2$CeRh$_4$Sn$_{13}$ monotonically decrease over the whole temperature range (Fig. \ref{fig:Fig_Waterfalls} (b)). The transition for the Ce rich material is clearly seen in Rietveld refinement as a gradual increase of ADPs (Fig. \ref{fig:Fig_Biso} (a)), which is not present on a Ca rich side (b). It is impossible to determine only from the diffraction data, if the origin of the increase of ADPs is static (phase transition) or dynamic (rattling of the central Sn1) and complementary studies are under way. There are also no signatures of the low temperature transition $T_D$ seen in resistance and susceptibility data. At this moment, it suggests that transition at $T_D$ is rather confined to electronic and not the crystal structure. On the other hand, the high temperature transition (with onset around 250 K and completed around $T_{HT}$ = 310~K), is most likely connected to an appearance of 2a~x~2a~x~2a superstructure observed in similar materials \cite{Bordet1991, Goh2015, Mazzone2015, Oswald2017, Lue2016, Otomo2013} but is below the detection limit for the Supernova setup. We can estimate that for CaCe$_2$Rh$_4$Sn$_{13}$ the $T_{HT}$ lies between 310~K and 320~K.

\subsubsection{Ca-La series}
In the Ca-La series a possible high temperature transition was observed around 250 K for Ca$_{2.8}$La$_{0.2}$Rh$_4$Sn$_{13}$ (Fig. \ref{fig:Fig_Waterfalls} (d)) but not for CaLa$_2$Rh$_4$Sn$_{13}$ (Fig. \ref{fig:Fig_Waterfalls} (c)).
The refinement of the temperature dependence of ADPs for Ca$_{2.8}$La$_{0.2}$Rh$_4$Sn$_{13}$ (Fig. \ref{fig:Fig_Biso} (d)) revealed 4 characteristic regions: (1) up to 140 K where ADPs anomalously decrease, (2) between 140 K and 250 K where they increase as expected from increased thermal fluctuations, (3) a sharp decrease around 250 K, which points towards sharp decrease of the disorder and (4) again a small step around 300 K. At this moment the increase of ADPs between 140 K and 250 K can be attributed to an increase of a static disorder which contributes to atomic displacements equally with the thermal noise. This disorder is eliminated by an onset of a possible transition at 250~K, which is completed at around 300~K. Similarly to the CaCe$_2$Rh$_4$Sn$_{13}$, this increase is most likely caused by a structural transition which is below the detection limit of the current Supernova setup. 

\begin{figure}[h!]
\includegraphics[width=0.48\textwidth]{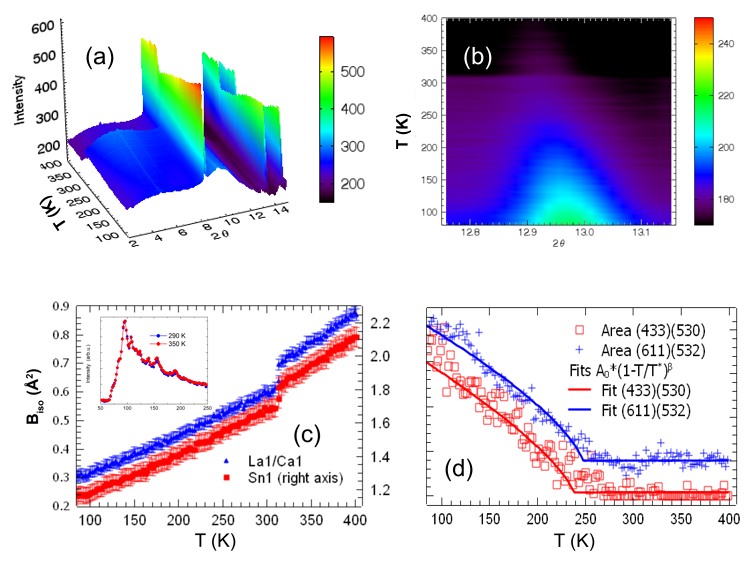}
\caption{\label{fig:Fig_Ca02}
Low section of synchrotron PXRD diffractograms for Ca$_{0.2}$La$_{2.8}$Rh$_4$Sn$_{13}$ (a) shows gradual variation of intensities and (b) appearance of very low overlapping 2a~x~2a~x~2a superstructure peaks (532)(611). The variation of intensities is correlated with a sudden jump in $B_{iso}$ at $T_{HT}=$310~K (c), which does not affect Raman spectra (inset). On the other hand the superstructure reflections display critical behavior with $T^{*} \approx$ 250~K (d).
}
\end{figure}

In order to justify the result, we present preliminary analysis of synchrotron PXRD carried out on Ca$_{0.2}$La$_{2.8}$Rh$_4$Sn$_{13}$ (Fig. \ref{fig:Fig_Ca02}), which lies on the La rich side of the Ca-La series and is expected to undergo the $T^{*}$ transition.
\\The low angle section of the pattern (Fig. \ref{fig:Fig_Ca02} (a)) reveals three important features:
(1) a sudden drop of the background around 310~K, which is accompanied by an increase of peak intensity, which we attributed to a transfer of scattering intensity from a long range order represented by Bragg reflections to a short ranged component visible as diffuse scattering, (2) an appearance of 2a~x~2a~x~2a superstructure reflections (overlapping (532) and (611) in \ref{fig:Fig_Ca02} (b)), which decrease gradually with the temperature having a critical-like behavior below 250~K. An order parameter-like curve fitted for the stronger pair of reflections (Fig. \ref{fig:Fig_Ca02} (d)) gave better estimates of parameters: the $T^{*} = 247 \pm 19$~K and the critical exponent $\beta = 0.71 \pm 0.06$. We do not imply here that this transition is necessarily continuous but the quality of the fit and earlier reports on similar materials \cite{Goh2015,Cheung2016} strongly support this conclusion. It must be noted that the intensity of the reflection is proportional to the $|F_{HKL}^2|$, which in the first approximation is proportional to the square of the deviation from the ideal symmetry. This means that the critical exponent calculated from intensities is twice as large as the one potentially calculated from the displacements (in a similar way intensities of magnetic peaks are proportional to the square of magnetic moment). Therefore the actual critical exponent of the order parameter will be equal to $\beta^{*} = 0.35 \pm 0.03$.

At the end, a quick look at representative ADPs (Fig. \ref{fig:Fig_Ca02} (c)) shows that the change in intensities of the main reflections and the jump of the background at $T_{HT}$ = 310~K is connected to displacive disorder. An attempt was done to differentiate between static and dynamic (increased rattling) origins of the transition by carrying out Raman measurements, in the energy range suitable for phonon excitations, at room temperature (below $T_{HT}$ but above $T^{*}$) and 350~K (above $T_{HT}$) (Fig. \ref{fig:Fig_Ca02} (c) inset). For better comparison a baseline was subtracted from both spectra and their respective maxima were normalized to 1. No significant difference in intensities is observed while crossing the 310~K boundary, therefore there is no evidence of sudden changes in vibrational/phonon modes. More plausible explanation of this transition is an appearance of a static disorder, which leads to static long range ordered superstructure at lower temperatures.\\
The possible presence of an additional high temperature transition $T_{HT}$, which is associated with a sudden jump in ADPs and a decrease in a background seems to be a precursor to the long range order developed at $T^{*}$ has not been previously reported in similar systems. Our data shows that it is present in samples with $x_{La} = 0.2$ (Fig. \ref{fig:Fig_Biso} (d)), $x_{La} = 2.8$ (Fig. \ref{fig:Fig_Ca02} (c)) and $x_{Ce} = 2.0$ (Fig. \ref{fig:Fig_Biso} (a)) and is most likely connected with the appearance of short range correlations between atomic displacements. A striking feature for the Ca-La series is that it is visible for the border compositions $x_{La} = 0.2$ and $x_{La} = 2.8$ but not for $x_{La} = 2.0$~ \cite{CommentPZ2}.

\begin{figure}[h!]
\includegraphics[width=0.48\textwidth]{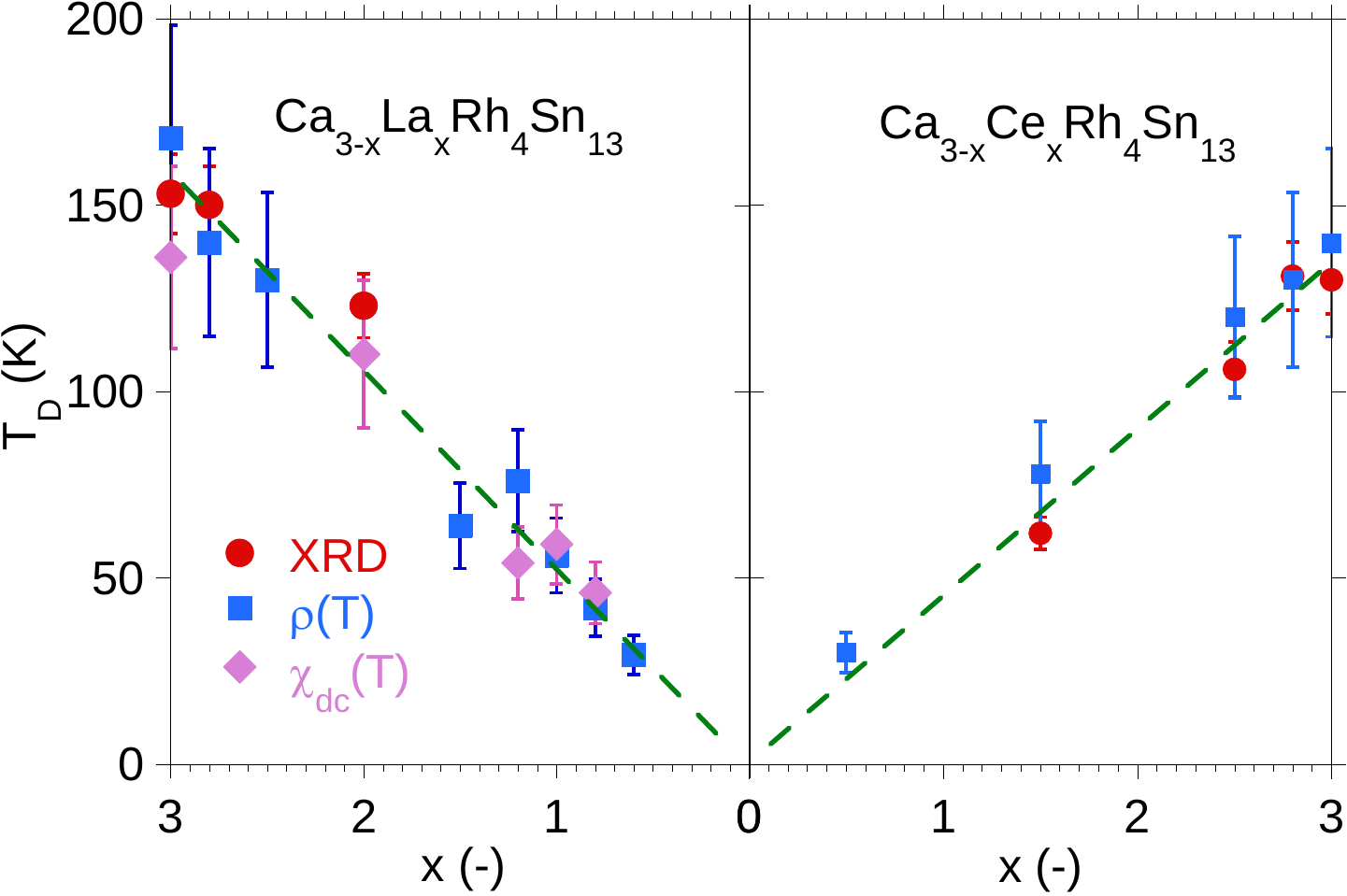}
\caption{\label{fig:Fig-Distortion_La-Ce}
Phase diagram showing the effect of lanthanum (Ca$_{3-x}$La$_{x}$Rh$_4$Sn$_{13}$) and cerium (Ca$_{3-x}$Ce$_{x}$Rh$_4$Sn$_{13}$) doping on temperature $T_{D}$, which represents  the change of the lattice dynamics of the respective alloys $x$ (see the text, also Ref. \onlinecite{Hu2017}). The red points are the temperatures $T_{D}$, determined for each sample $x$
from the $d$ vs $T$  plot in $log-log$ scale.  In such $log-log$ presentation of the $d(T)$ data, $T_{D}$ represents a clear kink between two linear dependencies in $d(T)$ at temperatures  $T>T_{D}$ and $T<T_{D}$, respectively. 
The similar kink at $T_{D}$ shows  dc magnetic susceptibility $\chi_{dc}(T)$ (pink diamonds), and 
inverse electrical resistivity $1/\rho (T)$   (blue square points),  when the both quantities are presented in $log-log$ scale.
}
\end{figure}

One can ask if the transition T$_D$ is somehow reflected in the structural properties. Due to limited angular resolution of Supernova scans, no subtle transitions could be detected in slopes of thermal expansion of lattice parameters (see the Supplementary Information) and detailed studies are under way. For the purpose of this paper we have repeated temperature scans with higher angular resolution over the angular range of (622) reflection. Interplanar d-spacing calculated in this way was plotted on a $log-log$ scale in a way similar to the resistance data (see Supplementary Information). The plot revealed two possible inflection points with temperatures equivalent to $T_D$ and $T^{*}$. At present we can only associate the lower transition with the $T_D$ since it perfectly correlates with temperatures observed in resistance and susceptibility data in both La and Ce dopes series (Fig. \ref{fig:Fig-Distortion_La-Ce})

\section{conclusions}
The comprehensive research on La and Ce substituted Ca$_{3}$Rh$_4$Sn$_{13}$ leads to several important observations, which are helpful for a better understanding of  the nature of its superconductivity. The basic problem of our studies is an atomic disorder, which plays a crucial role in the strongly correlated materials. It was documented theoretically for this class of compounds that in the critical regimes near the quantum critical point a system is at the threshold of an instability, and disorder as perturbation, can cause significant macroscopic effects. Due to structural disorder of the doped Ca$_{3}$Rh$_4$Sn$_{13}$, one has to distinguish two behaviors. First, the atomic disorder leads to significant decrease in $T_c$ of Ca$_{3}$Rh$_4$Sn$_{13}$ upon quenching. This $T_c$ decrease can  easily be explained from the BCS equation \cite{Bardeen1957} $T_c=\theta_De^{-1/N(\epsilon_F)U}$ as consequence of local stress due to disorder, which significantly changes $N(\epsilon_F)$ value, and thus changes the expression $N(\epsilon_F)U\sim \frac{\lambda - \mu^{\star}}{1+\lambda}$~ \cite{Seiden1969}. Secondly, the doping of Ca$_{3}$Rh$_4$Sn$_{13}$  generates the nanoscale electron disorder in the bulk sample, leading to an inhomogeneous superconductivity state with an enhanced critical temperature $T^{\star}_c >T_c$. The $T-x$ diagram (Fig. \ref{fig:Fig_DIAGRAM_T-x-sum}) and the entropy $S(x)_T$ isotherms (Fig. \ref{fig:Fig_S-vs-x_La-Ce-T-const}) well document the relation between degree of an atomic disorder and separation of two superconducting phases, $T^{\star}_c$ and $T_c$, in the $T$-scale. We interpret the effect of the high temperature $T^{\star}_c$ phase as a result of its larger lattice stiffening with respect to the bulk superconducting phase $T_c$. Based on the  Eliashberg theory of BCS superconductivity and the band structure calculations, we propose a phenomenological model, which qualitatively describes the resistivity data obtained under high pressure. We also demonstrated that the $T-x$ and $S(x)_T$ data well correlate with the appearance of superstructure at $T_{HT}\sim 310$ K. This high-temperature second-order phase transition is clearly observed for the Ca$_{3-x}$La$_x$Rh$_4$Sn$_{13}$ and Ca$_{3-x}$Ce$_x$Rh$_4$Sn$_{13}$ alloys having 
 relatively low disorder due to doping, i.e., located in $T-x$ diagram near the parent compounds  Ca$_{3}$Rh$_4$Sn$_{13}$,  La$_{3}$Rh$_4$Sn$_{13}$, or  Ce$_{3}$Rh$_4$Sn$_{13}$, while for $x$-range, where the isotherms $S(x)_T$ show a maxima that correspond to the largest separation between the superconducting phases $T^{\star}_c$ and $T_c$ the superstructure effect is missing. This means that strong atomic disorder generated by doping is not significant enough so that structural changes can be observed
 at $T_{HT}$ within an experimental detection limit.
 
In such  dirty Ca$_{3-x}$La$_x$Rh$_4$Sn$_{13}$ and Ca$_{3-x}$Ce$_x$Rh$_4$Sn$_{13}$ superconductors, where the mean free path is much smaller than the coherence length, the upper critical field $H_{c2}(T)$ is usually interpreted within the Werthamer-Helfand-Hohenberg  or Maki--de Gennes  theoretical model. 
 This  model, however, does not well fit the  $H_{c2}$ vs. $T_c$ data, and does not interpret the observed   positive curvature of $H_{c2}$ close to $T_c$. One of the possible scenarios of the observed $H_{c2}(T)$ dependencies could be a two-band nature of superconductivity in these systems. Previously, intriguing evidence for multiband  effects was also documented in structurally similar filled skutterudite LaRu$_4$As$_{12}$~\cite{Cichorek2012} and LaOs$_4$As$_{12}$~\cite{Cichorek2015} compounds. However, it is also possible that the shape of $H_{c2}(T)$ stems from the presence of nanoscopic inhomogeneities of the superconducting state.

\section{acknowledgments}
The research was supported by National Science Centre (NCN) on the basis of Decision No. DEC-2012/07/B/ST3/03027.
M.M.M. acknowledges support by NCN under grant DEC-2013/11/B/ST3/00824. P.W. acknowledges support by NCN under grant DEC-2015/17/N/ST3/02361.
High-pressure research at the University of California, San Diego, was supported by the National Nuclear Security Administration under the Stewardship Science Academic Alliance program through the U. S. Department of Energy under Grant Number DE-NA0002909. One of us (A.\'S.) is grateful for the hospitality at the University of California, San Diego (UCSD). We are grateful to Dr D. Chernyshov and Dr W. van Beek from ESRF (Swiss Norwegian Beamlines) for help with diffraction measurements.

\newpage
\includegraphics[width=\textwidth]{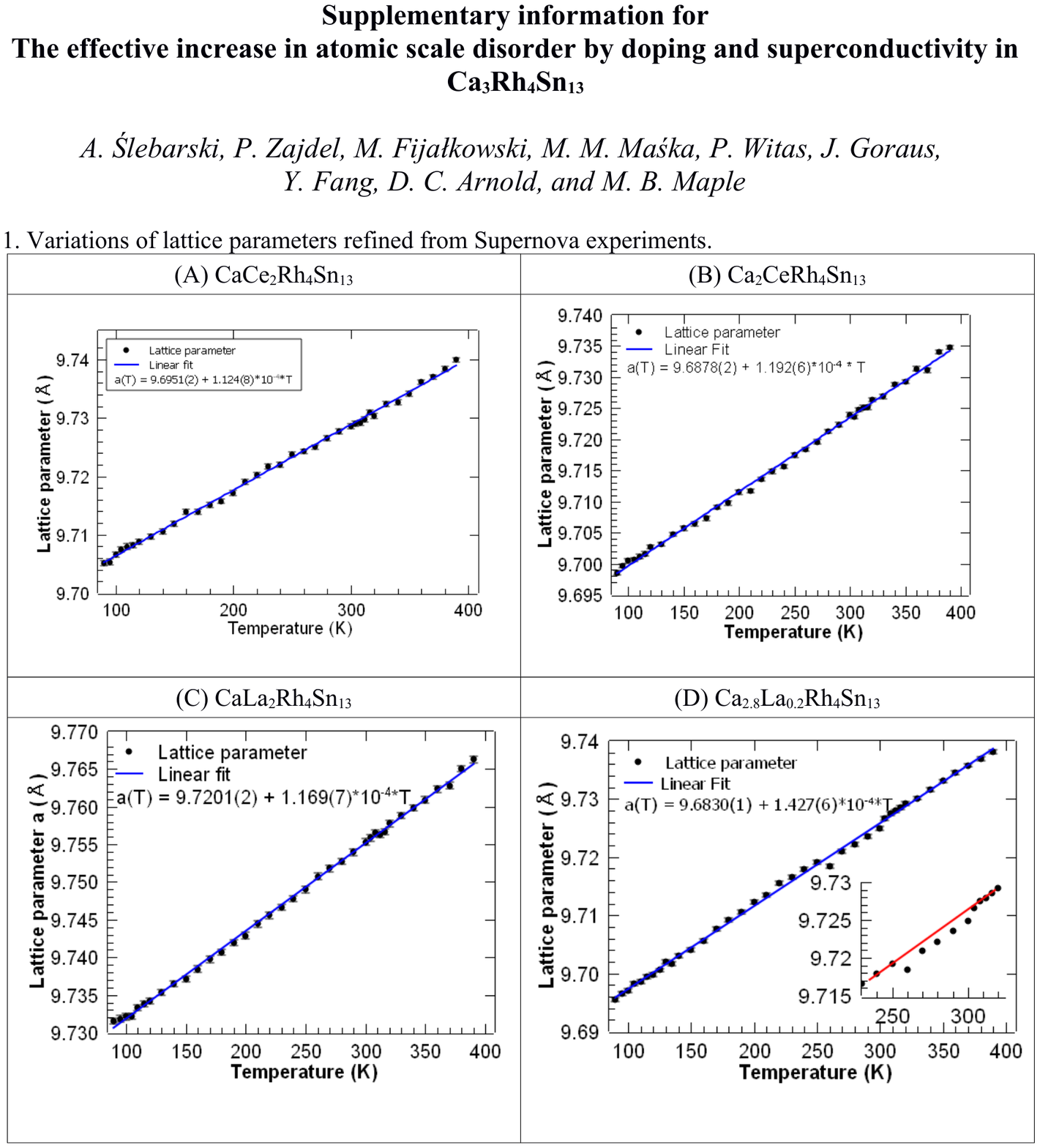}
\clearpage
\includegraphics[width=\textwidth]{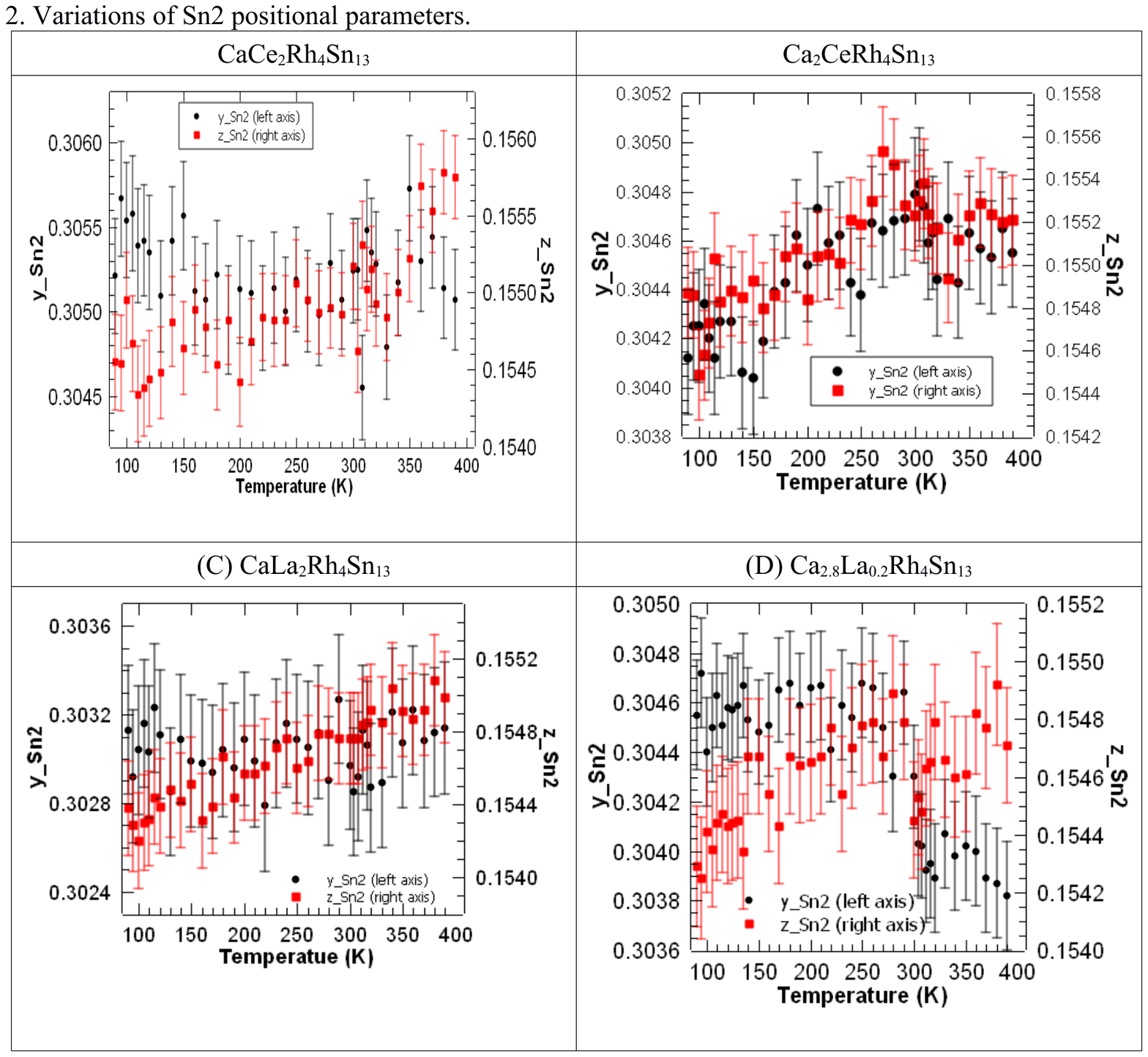}
\clearpage
\hspace*{-9.3cm}\includegraphics[width=\textwidth]{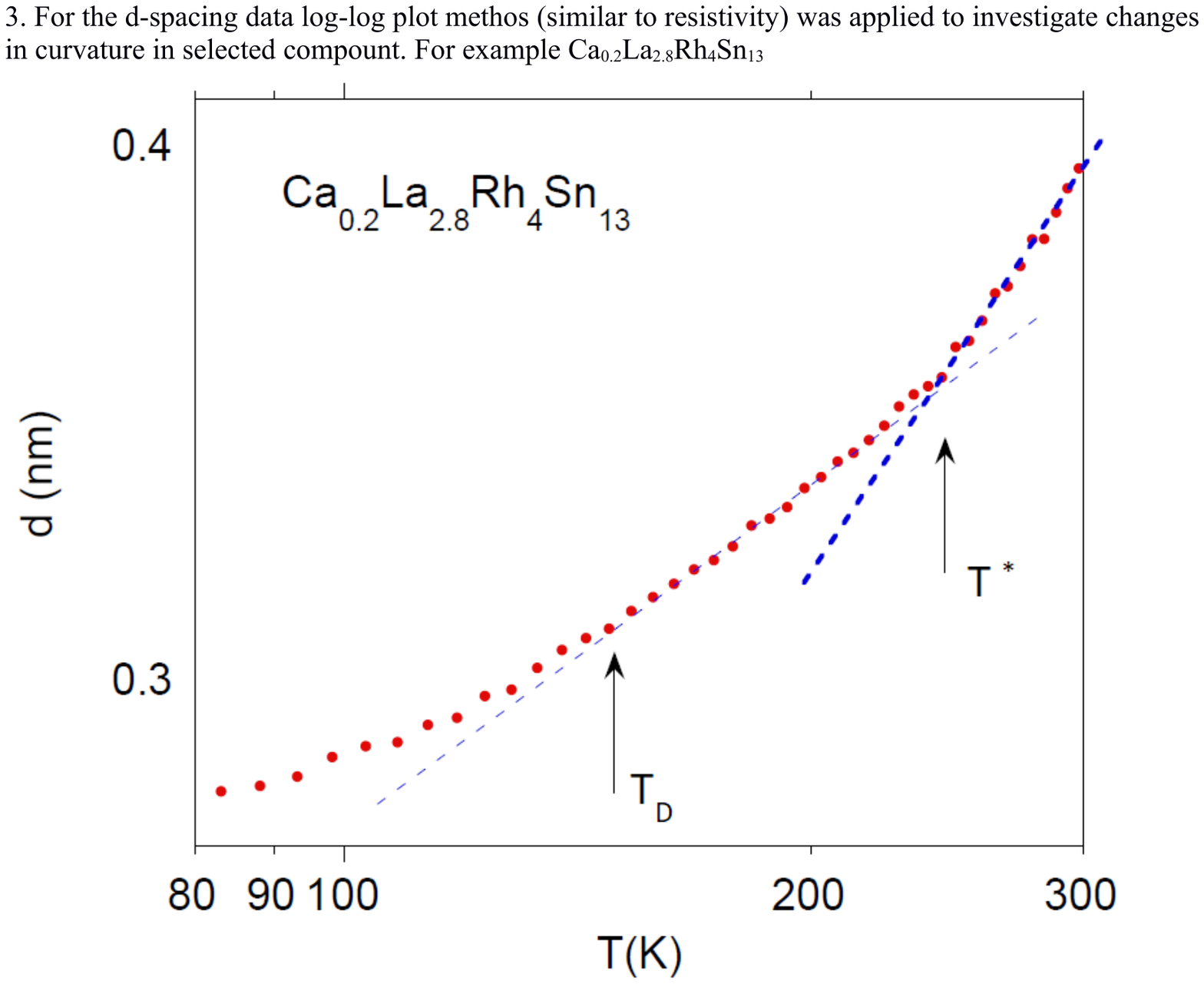}


\begin{thebibliography}{99}
\bibitem{comment1}
The long range magnetic order has not been observed for $x$ components of  Ca$_{3-x}$Ce$_x$Rh$_4$Sn$_{13}$ in temperatures $T>0.4$ K, however, presence of short-range magnetic correlations with spin-glass-like behavior are confirmed by ac magnetic susceptibility data, magnetization $M$ vs $B$, and specific heat.
\bibitem{CommentPZ1}
Refinements of measurements carried out with Cu K wavelength (1.54 {{\AA}}) require an absorption correction, which is very difficult to estimate for the applied geometry and results in negative values of ADPs. An attempt was made to apply a moderate cylindrical absorption correction (${\mu r}$~=~3.0) but it was not sufficient to fully correct the data. In this case, the values of ADP have wrong absolute values but their mutual relations and general trends are preserved.
\bibitem{CommentPZ2}
The compositions of the Ca rich sample was additionally checked by SEM after the XRD.
\end{thebibliography}
\end{document}